\documentclass[10pt,a4paper]{article}
\usepackage[utf8]{inputenc}
\usepackage{a4wide}
\usepackage{graphicx}
\usepackage{hyperref}

\usepackage{amsmath}
\usepackage{verbatim} 

\begin{document}

\begin{titlepage}
\begin{flushright}
\end{flushright}
\begin{flushright}
LU TP 16-43\\
November 2016
\end{flushright}
\vfill
\begin{center}
{\Large\bf 	An Analytic Approach to Sunset Diagrams in Chiral Perturbation Theory: Theory and Practice}
\vfill
{\bf B. Ananthanarayan$^a$, Johan Bijnens$^b$, Shayan Ghosh$^a$, Aditya Hebbar$^{a,c}$}\\[1cm]
{$^a$ Centre for High Energy Physics, Indian Institute of Science, \\
Bangalore-560012, Karnataka, India}\\[0.5cm]
{$^b$Department of Astronomy and Theoretical Physics, Lund University,\\
S\"olvegatan 14A, SE 223-62 Lund, Sweden} \\[0.5cm]
{$^c$Department of Physics and Astronomy, University of Delaware,\\
Newark, DE 19716, USA\footnote{Present Address}} \\
\end{center}
\vfill
\begin{abstract}
We demonstrate the use of several code implementations of the Mellin-Barnes method available in the public domain to derive analytic expressions for the sunset diagrams that arise in the two-loop contribution to the pion mass and decay constant in three-flavoured chiral perturbation theory. We also provide results for all possible two-mass configurations of the sunset integral, and derive a new one-dimensional integral representation for the one mass sunset integral with arbitrary external momentum. Thoroughly annotated Mathematica notebooks are provided as ancillary files, which may serve as pedagogical supplements to the methods described in this paper.

\end{abstract}
\vfill
\vfill
\end{titlepage}

\section{Introduction}

Chiral perturbation theory is a low energy effective field theory of the strong interaction. The work \cite{Kaiser:2007kf} presents analytic expressions for the two-loop contribution to the pion mass and decay constant in SU(3) chiral perturbation theory with suitable expansions in powers of $m_{\pi}^2$. In an upcoming work \cite{ABG}, we will present analogous expressions for the pion decay constant. Work is also underway to find similar simple analytic representations for the kaon and eta mass and decay constants to two loops.

Due to the Goldstone nature of the particles involved, scalar, tensor and derivatives of sunset diagrams appear in these calculations, with various mass configurations and with up to three distinct masses. Much work has been done on sunset diagrams (an incomplete list is given in references \cite{Amoros:1999dp}-\cite{Czyz:2002re}), and a variety of analytic results exist in the literature for the one-and two-mass scale configurations \cite{Amoros:1999dp, Berends:1997vk, Gasser:1998qt, Davydychev:1992mt, Adams:2015gva, Post:1996gg, Kniehl:2005bc, Martin:2003qz, Czyz:2002re}. Papers directly relevant to this work are the following. In \cite{Berends:1997vk}, analytic results have been given for the master integrals at the pseudothreshold $ s = (m_1+m_2-m_3)^2 $ and threshold $ s = (m_1+m_2+m_3)^2 $, the former of which may be used to obtain the single, and many of the double, mass scale analytic expressions. Gasser and Saino \cite{Gasser:1998qt} use integral representations to give results in closed form for several basic two-loop integrals appearing in ChPT, including the sunset, with one mass-scale. For unequal masses, fully analytic results are given in \cite{Adams:2015gva} in terms of newly defined elliptic generalizations of the Clausen and Glaisher functions, but the application of methods or approximation schemes that give the three mass scale sunsets as expansions in powers of the mass ratio allow for a more transparent interpretation of the results being considered. In \cite{Berends:1993ee}, just such an expansion is given for the most general sunset integral in terms of Lauricella functions. However, none of the series presented in \cite{Berends:1993ee} converge for the physical values of the meson masses.

The interest in analytic or semi-analytic expressions arises from the desire to make as direct a contact as possible with results in lattice field theories. Recent advances in lattice QCD now allow for quark masses in these theories to be varied independently, allowing for realistic quark masses. The availability of analytic results for pseudo-scalar masses and decay constants, for example, would allow for easy and computationally efficient comparison with lattice results.

Aside from the derivation of analytic expressions for the pseudo-scalar meson masses and decay constants to two-loops, the application of sunset diagrams to chiral perturbation theory is also of general interest. In this context,  sunset diagrams have been studied quite early (\cite{Post:1996gg}), where not only the single mass scale sunset (which appears in SU(2) chiral perturbation theory) is considered, but also the cases with more than one mass scale which are common in the SU(3) theory. In SU(3) chiral perturbation theory, the sunset is the simplest diagram that appears at two loops, and a careful study of it paves the way for the study of the other diagrams that appear at this order (i.e.  vertices, boxes and acnodes). The work  \cite{Gasser:1998qt} gives a terse but comprehensive summary of results. Another possible use of the sunsets is to expand them out using methods such as expansion in regions \cite{Kaiser:2006uv}, and then use this to reduce the SU(3) low energy constants to the SU(2) ones. The process of relating the SU(3) to SU(2) low-energy-constants has been done using an alternative method in \cite{Gasser:2010zz} but it has not yet been done for the full set of low-energy-constants at next-to-next-to-leading order. It must be noted in the context of \cite{Kaiser:2006uv} that the sunset technology is also important when considering vertices, as many of the latter get related to the sunsets when using, for example, the method of expansion by regions.

In this paper, we use the Mellin-Barnes method to derive results for all the single and double mass scale integrals. It has been shown in \cite{Friot:2005cu} that the Mellin-Barnes method is an efficient one for obtaining expansions in ratios of two mass scales should they appear in Feynman diagrams in general. This work therefore serves as an independent verification of the existing results in the literature. The Mellin-Barnes method is also an appropriate tool for chiral perturbation theory applications as it ab initio allows us to express the integrals as expansions in mass ratios.

A further reason for Mellin-Barnes as our tool of choice is the availability of powerful public computer packages in this approach. The availability of such  codes has made such a study of sunsets (and two-loop diagrams in general) in chiral perturbation theory much more accessible.  The Mathematica based package \texttt{Tarcer} \cite{Mertig:1998vk} applies the results of Tarasov's work \cite{Tarasov:1997kx} to recursively reduce all sunset diagrams to the master integrals. Several packages \cite{Gluza:2007rt, Smirnov:2009up, Czakon:2005rk} have automatized many aspects of the application of Mellin-Barnes methods to Feynman integrals. The sunsets appearing in chiral perturbation theory have been implemented numerically in the package \texttt{Chiron}  \cite{Bijnens:2014gsa} using the methods of \cite{Amoros:1999dp}. One of the goals of the present work is to improve on this implementation. In addition, there are two other packages \texttt{BOKASUM} \cite{Caffo:2008aw} and \texttt{TSIL} \cite{Martin:2005qm} that can be used to numerically calculate sunset integrals.

We present along with this paper several Mathematica notebooks (lodged as ancillary files along with the arXiv submission) which contain the details of our calculations, as well as a demonstration of how to apply the above packages to the calculation of sunset integrals. The notebooks are thoroughly annotated, and can be used in a stand-alone capacity, or in conjunction with this note. These may also serve as pedagogical introductions to the analytic evaluation of sunset diagrams.

The primary goal of this paper is to show the use of the packages of \cite{Mertig:1998vk, Gluza:2007rt, Smirnov:2009up, Czakon:2005rk, Kosower} but the results as presented here have been checked in a number of other ways as well. The relations from \cite{Mertig:1998vk} have been implemented independently using \texttt{FORM} \cite{Vermaseren:2000nd}. The expansions around $s=0$ were also derived using the methods of \cite{Amoros:1999dp, Post:1996gg} and numerical results have been compared with the results from analytical expressions of \cite{Berends:1997vk, Martin:2003qz, Czyz:2002re}.

This paper is organized as follows. In Section \ref{secTwoLoops}, we give the five different sunset configurations that will be explicitly considered in this work, and show from where they arise. In Section \ref{secSunsets} we give an overview of the sunset integrals, their divergences, and their renormalization in chiral perturbation theory. In Section \ref{secMB}, we briefly discuss the Mellin-Barnes method of evaluating Feynman integrals. In Section \ref{secSpecialSunsets}, we demonstrate the use of the package \texttt{Tarcer} \cite{Mertig:1998vk} to reduce the tensor and derivatives of the sunsets to master integrals. In Section \ref{sec1Mass}, we explain the use of the packages \cite{Gluza:2007rt, Smirnov:2009up, Czakon:2005rk, Kosower} to derive the results for the one-mass scale master integral. We also explain how the \texttt{Tarcer} package \cite{Mertig:1998vk} alone can be used to derive this result. In Section \ref{sec2Mass}, we describe briefly the two different categories of two-mass scale sunset diagrams and their evaluation, and present a complete set of results in Appendix \ref{sec2MassNPT}. In Section \ref{sec3Mass}, we explain how three mass scale sunsets can be handled either by means of an expansion in the external momentum, or by a more sophisticated application of the Mellin-Barnes method. In Section \ref{sec1DRep}, we present a one-dimensional integral representation of an important configuration that arises in the SU(2) chiral perturbation theory, and in Section \ref{secNumerics} with a discussion of some numerical issues of the new results presented herein. We conclude in Section \ref{secConcl} with a discussion of the relevance and limitations of this work, and possible future work in this field. In Appendix \ref{secPublicCodes}, we give a brief description of all the public codes used in this work, and in Appendix \ref{secNotation}, we present a dictionary that allows for an easy translation between the definition used in this work for the sunset and other integrals, and those used in the various programs and papers. In Appendix \ref{secAncFiles}, we list the ancillary files provided with this paper.

\section{The Meson Masses and Decay Constants to Two Loops \label{secTwoLoops}}

Expressions for the pseudoscalar meson masses and decay constants in two loop chiral perturbation theory are given in \cite{Amoros:1999dp}. As a concrete example, the pion mass is given by:
\begin{align}
	m^2_{\pi} = m^2_{0 \pi} + \left( m^2_{\pi} \right)^{(4)} + \left( m^2_{\pi} \right)^{(6)}_{CT} + \left( m^2_{\pi} \right)^{(6)}_{loops} + \mathcal{O} \left(p^8 \right)
\end{align}
where  $m^2_{0 \pi}$ is the bare mass, $\left( m^2_{\pi} \right)^{(4)}$ is the one-loop contribution, $\left( m^2_{\pi} \right)^{(6)}_{CT}$ is the two-loop model-dependent counterterm contribution, and $\left( m^2_{\pi} \right)^{(6)}_{loops}$ is the chiral loop contribution.

It is in this last term that the sunset integrals appear:
\begin{align}
	F^4_{\pi} \left( m^2_{\pi} \right)^{(6)}_{loops} = ... & + 5/6 H\left( m_{\pi}^2, m_{\pi}^2, m_{\pi}^2; m_{\pi}^2 \right) m_{\pi}^4 - 5/8 H\left( m_{\pi}^2, m_K^2, m_K^2; m_{\pi}^2 \right) m_{\pi}^4 \nonumber \\
	& + 1/18 H\left( m_{\pi}^2, m_{\eta}^2, m_{\eta}^2; m_{\pi}^2 \right) m_{\pi}^4 + H\left( m_{K}^2, m_{\pi}^2, m_{K}^2; m_{\pi}^2 \right) m_{\pi}^2 m_{K}^2  \nonumber \\
	& - 5/6 H\left( m_{K}^2, m_{K}^2, m_{\eta}^2; m_{\pi}^2 \right) m_{\pi}^4
	  - 1/8 H\left( m_{\eta}^2, m_{K}^2, m_{K}^2; m_{\pi}^2 \right) m_{\pi}^4  \nonumber \\
	& + 1/2 H\left( m_{\eta}^2, m_{K}^2, m_{K}^2; m_{\pi}^2 \right) m_{\pi}^2 m_{K}^2
	  + H_{1}\left( m_{\pi}^2, m_{K}^2, m_{K}^2; m_{\pi}^2 \right) m_{\pi}^4  \nonumber \\
	& + 2 H_{1}\left( m_{K}^2, m_{K}^2, m_{\eta}^2; m_{\pi}^2 \right) m_{\pi}^4
	+ 3 H_{21}\left( m_{\pi}^2, m_{\pi}^2, m_{\pi}^2; m_{\pi}^2 \right) m_{\pi}^4 \nonumber \\
	& - 3/8 H_{21}\left( m_{\pi}^2, m_{K}^2, m_{K}^2; m_{\pi}^2 \right) m_{\pi}^4
	  + 3 H_{21}\left( m_{K}^2, m_{\pi}^2, m_{K}^2; m_{\pi}^2 \right) m_{\pi}^4  \nonumber \\
	& + 9/8 H_{21}\left( m_{\eta}^2, m_{K}^2, m_{K}^2; m_{\pi}^2 \right) m_{\pi}^4 \label{pionmass}
\end{align}

The $H$ in the above expression refer to the scalar sunset integral $H^d_{\{1,1,1\}}$ as defined in Eq.(\ref{sunsetdef}) of Section \ref{secSunsets}, where the first three arguments pertain to the masses entering the propagators, and the last is the square of the energy entering the loop. The $H_1$ and $H_{21}$ are the scalar integrals that make up the Passarino-Veltman decomposition of vector and tensor sunsets, and are defined precisely in Eq.(\ref{H1}) and Eq.(\ref{H21}) respectively.

In the case of the meson decay constants, in addition to the variety of sunset integrals appearing above, also appear derivatives of the sunsets (i.e.$H'$, $H'_{1}$ and $H'_{21}$). The work of finding an analytic expression for the pion mass (as well as the other pseudoscalar meson masses and decay constants) reduces to analytically evaluating these sunset integrals.

In the subsequent sections of this paper, we explain how to analytically evaluate each of the different types of integrals appearing in expressions such as Eq.(\ref{pionmass}) above. In particular, we show in detail how to evaluate the following integrals as representative of the different types of integrals and the different types of mass configurations that may appear in expressions for the pseudoscalar masses and decay constants:

\begin{table}
\centering
\begin{tabular}{|c|c|}
\hline 
\rule[-1ex]{0pt}{2.5ex} \textbf{Integral} & \textbf{Characteristic} \\ 
\hline 
\rule[-1ex]{0pt}{2.5ex} $H\left( m_{\pi}^2, m_{\pi}^2, m_{\pi}^2; m_{\pi}^2 \right) $ & One mass scale\\ 
\hline 
\rule[-1ex]{0pt}{2.5ex} $H\left( m_{\pi}^2, m_K^2, m_K^2; m_{\pi}^2 \right)$ & Two mass scales \\ 
\hline 
\rule[-1ex]{0pt}{2.5ex} $H\left( m_{\eta}^2, m_{K}^2, m_{K}^2; m_{\pi}^2 \right)$ & Three mass scales with smallest parameter as external momentum\\ 
\hline 
\rule[-1ex]{0pt}{2.5ex} $H\left( m_{K}^2, m_{K}^2, m_{\pi}^2; m_{\eta}^2 \right)$ & Three mass scales with an internal mass as smallest parameter\\
\hline 
\rule[-1ex]{0pt}{2.5ex} $H'_{21}\left( m_{\pi}^2, m_{K}^2, m_{K}^2; m_{\pi}^2 \right)$ & Tensor sunset derivative \\ 
\hline 
\end{tabular}
\caption{Examples of sunset integrals and mass configurations that appear in expressions for the meson masses and decay constants at two-loops}
\end{table}

The evaluation of all these integrals requires writing them in terms of master integrals, and then analytically evaluating the master integrals. This is explained in greater detail in the next section. The analytic evaluation of the master integrals can be done using a variety of methods, and many of these have previously been used to derive the plethora of results that exist in the literature. In this paper, we use the Mellin-Barnes approach, which appears to be the most efficient method by which to evaluate the three mass scale integrals, such as $H\left( m_{K}^2, m_{K}^2, m_{\pi}^2; m_{\eta}^2 \right)$ that appears in the expressions for eta mass and decay constant.

The integrals given in the table above are all amenable to a Mellin-Barnes treatment. However, for $H\left( m_{\eta}^2, m_{K}^2, m_{K}^2; m_{\pi}^2 \right)$, we instead take an expansion in the external momentum $s=m_{\pi}^2$, as it provides a result that is as accurate as a Mellin-Barnes expansion (to the same order) but that is much easier to calculate. A similar expansion cannot be done for $H\left(m_{K}^2, m_{K}^2, m_{\pi}^2; m_{\eta}^2 \right)$ in either the external momentum $s=m_{\eta}^2$ due to poor convergence, or in $m_{\pi}^2$ as it gives rise to an infrared divergence.

\section{Sunset Integrals \label{secSunsets}}

\begin{figure}[hbtp]
\centering
\includegraphics[scale=0.6]{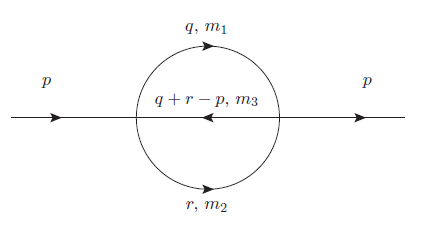}
\caption{Sunset diagram}
\label{sunset}
\end{figure}

The sunset integral, shown in Figure \ref{sunset}, is defined as:
\begin{align}
H_{\{\alpha,\beta,\gamma\}}^d \{m_1,m_2,m_3; s=p^2\} = \frac{1}{i^2} \int \frac{d^dq}{(2\pi)^d} \frac{d^dr}{(2\pi)^d} \frac{1}{[q^2-m_1^2]^{\alpha} [r^2-m_2^2]^{\beta} [(q+r-p)^2-m_3^2]^{\gamma}} \label{sunsetdef}
\end{align}

Vector and tensor sunset integrals have four-momenta, such as $q_{\mu}$ or $q_{\mu}q_{\nu}$, sitting in the numerator. Two tensor integrals that appear in the calculation of meson masses and decay constants in chiral perturbation theory are:
\begin{align}
	H_{\mu}^d \{m_1,m_2,m_3; p^2\} = \frac{1}{i^2} \int \frac{d^dq}{(2\pi)^d} \frac{d^dr}{(2\pi)^d} \frac{q_{\mu}}{[q^2-m_1^2] [r^2-m_2^2][(q+r-p)^2-m_3^2] } \nonumber \\
	\nonumber \\
	H_{\mu \nu}^d \{m_1,m_2,m_3; p^2\} = \frac{1}{i^2} \int \frac{d^dq}{(2\pi)^d} \frac{d^dr}{(2\pi)^d} \frac{q_{\mu} q_{\nu}}{[q^2-m_1^2] [r^2-m_2^2] [(q+r-p)^2-m_3^2] }
\end{align}

These may be decomposed into linear combinations of scalar integrals via the Passarino-Veltman decomposition as:
\begin{align}
	& H_{\mu}^d = p_{\mu} H_{1} \nonumber \\
	& H_{\mu \nu}^d = p_{\mu} p_{\nu} H_{21} + g_{\mu \nu} H_{22}
\end{align}

To obtain the scalar integral $H_{1}$, we take the scalar product of $H_{\mu}^d$ with $p^{\mu}$:
\begin{align}
	H_{1} & = \frac{1}{p^2} \frac{1}{i^2} \int \frac{d^dq}{(2\pi)^d} \frac{d^dr}{(2\pi)^d} \frac{p.q}{[q^2-m_1^2] [r^2-m_2^2] [(q+r-p)^2-m_3^2] } \equiv \frac{1}{p^2} \langle \langle q.p \rangle \rangle \label{H1}
\end{align}
where we define $\langle \langle X \rangle \rangle$ as the scalar sunset diagram with unit powers of the propagators, and with $X$ in the numerator. 

Similarly, $H_{21}$ may be expressed as:
\begin{align}
	H_{21} = \frac{\langle \langle (q.p)^2 \rangle \rangle d - \langle \langle q^2 \rangle \rangle p^2}{p^4(d-1)} \label{H21}
\end{align}

In \cite{Tarasov:1997kx} Tarasov has shown by using the method of integration by parts that all sunset diagrams, including those of higher than $d$ dimensions, may be rewritten as linear combinations of a set of four master integrals and bilinears of one-loop tadpole integrals. These basic integrals are $H^d_{\{1,1,1\}}, H^d_{\{2,1,1\}}, H^d_{\{1,2,1\}}, H^d_{\{1,1,2\}}$ and the one-loop tadpole integral:
\begin{align}
	A^d\{m\} = \frac{1}{i} \int \frac{d^d q}{(2\pi)^d} \frac{1}{q^2 - m^2} = - \frac{m^{d-2}}{(4\pi)^{d/2}} \Gamma \left( 1 - \frac{d}{2} \right) \label{tadpoledef}
\end{align}

Application of Tarasov's relations becomes crucial when evaluating another class of integrals that show up in chiral perturbation theory calculations, namely the derivatives of scalar and tensor sunsets (e.g. $H'_{\{1,1,1\}}, H'_{\{2,1,1\}}$). These may be evaluated by means of the following well-known formula relating derivatives and integrals in different dimensions \cite{Kaiser:2007kf,Tarasov:1997kx}:
\begin{align}
	\left( \frac{\partial}{\partial s} \right)^n H_{\{\alpha,\beta,\gamma\}}^d = (-1)^n ( 4 \pi )^{2n} \frac{\Gamma(\alpha+n) \Gamma(\beta+n) \Gamma(\gamma+n)}{\Gamma(\alpha) \Gamma(\beta)\Gamma(\gamma)} H_{\{\alpha+n,\beta+n,\gamma+n\}}^{d+2n} \label{DH}
\end{align}

The Mathematica package \texttt{Tarcer} \cite{Mertig:1998vk} automatizes the reduction of any sunset integral to the master integrals. Many results exist in the literature regarding these master integrals. One result that we use frequently in the subsequent sections is that of the two-mass scale master integral with zero external momentum. This is given in \cite{Davydychev:1992mt} as:
\begin{align}
	\left( 4 \pi \right)^4 & H^{\chi}_{\{1,1,1\}}\{M,M,m;0\} \nonumber \\	
	= & M^2 \bigg\{ \frac{x-4}{2} F \left[ x \right] - \frac{x}{2} \text{ ln}^2 \left[ x \right] + \left( 2 + x \right) \left[ \frac{\pi^2}{12} + \frac{3}{2} \right] \bigg\} \nonumber \\
	& - (\mu^2)^{-2\epsilon} \bigg\{ m^2 \log \left( \frac{m^2}{\mu^2} \right) \left[ 1 - \log \left( \frac{m^2}{\mu^2} \right) \right] + 2 M^2 \log \left( \frac{M^2}{\mu^2} \right) \left[ 1 - \log \left( \frac{M^2}{\mu^2} \right) \right] \bigg\} \nonumber \\	
	& + \frac{M^2}{2} \bigg\{ \bigg[ 2 + x \bigg] \frac{1}{\epsilon^2} + \bigg[ x \left( 1 - 2  \log \left( \frac{m^2}{\mu^2} \right) \right) + 2 \left( 1 - 2  \log \left( \frac{M^2}{\mu^2} \right) \right) \bigg] \frac{1}{\epsilon} \bigg\} + \mathcal{O}(\epsilon) \label{K111}
\end{align}
where 
\begin{align}
& x = m^2/M^2 \nonumber \\
& F(x) = \frac{1}{\sigma} \bigg[ 4 \text{Li}_2 \bigg( \frac{\sigma-1}{\sigma+1} \bigg) + \log^2 \bigg( \frac{1-\sigma}{1+\sigma} \bigg) + \frac{\pi^2}{3} \bigg] , \qquad \sigma = \sqrt{1-\frac{4}{x}}  
\end{align}

Eq.(\ref{K111}) above is the result for $H^{d}_{\{1,1,1\}}\{M,M,m;0\}$ to which the subtraction scheme normally used in chiral perturbation theory ($\overline{\text{MS}}_{\chi}$), which is a modified version of the $\overline{MS}$ scheme, has been applied. This is indicated by use of the index $\chi$ instead of $d$, and involves multiplying Eq.(\ref{sunsetdef}) by the factor $(\mu_{\chi}^2)^{4-d}$, where:
\begin{align}
	\mu_{\chi}^2 \equiv \mu^2 \frac{e^{\gamma_E-1}}{4\pi} \label{chiralMSbar}
\end{align}
In the remainder of this paper, unless explicitly stated, $H^{\chi}_{\{\alpha,\beta,\gamma\}}$ will be used to denote the finite part of the sunset integral evaluated using the $\overline{\text{MS}}_{\chi}$ scheme.

Analytic expressions for the divergent parts of the sunset master integrals have been derived in \cite{Czyz:2002re}, amongst other places. The following are the divergent parts of the master integrals in the $\overline{\text{MS}}_{\chi}$ scheme:
\begin{align}
	H_{\{1,1,1\}}^{\chi,div} & \{m_1,m_2,m_3;s\} = \frac{1}{512 \pi^4} \bigg\{ \left[ m_1^2+m_2^2+m_3^2 \right] \frac{1}{\epsilon^2} \nonumber \\
	& + \left[ m_1^2+m_3^2+m_3^2 - \frac{s}{2} - 2m_1^2 \log \left(\frac{m_1^2}{\mu ^2}\right) -  2m_2^2 \log \left(\frac{m_2^2}{\mu ^2}\right)- 2m_3^2 \log \left(\frac{m_3^2}{\mu ^2}\right) \right] \frac{1}{\epsilon} \bigg\} \nonumber \\
H_{\{2,1,1\}}^{\chi,div} & \{m_1,m_2,m_3;s\} = \frac{1}{512 \pi^4} \bigg\{ \frac{1}{\epsilon^2} - \left[ 1+ 2 \log \left(\frac{m_1^2}{\mu ^2}\right) \right] \frac{1}{\epsilon} \bigg\}
	\nonumber \\
H_{\{1,2,1\}}^{\chi,div} & \{m_1,m_2,m_3;s\} = \frac{1}{512 \pi^4} \bigg\{ \frac{1}{\epsilon^2} - \left[ 1+ 2 \log \left(\frac{m_2^2}{\mu ^2}\right) \right] \frac{1}{\epsilon} \bigg\}
	\nonumber \\
H_{\{1,1,2\}}^{\chi,div} & \{m_1,m_2,m_3;s\} = \frac{1}{512 \pi^4} \bigg\{ \frac{1}{\epsilon^2} - \left[ 1+ 2 \log \left(\frac{m_3^2}{\mu ^2}\right) \right] \frac{1}{\epsilon} \bigg\} \label{divParts}
\end{align}
Eq.(\ref{chiralMSbar}) may be reverse engineered and used in combination with Eq.(\ref{divParts}) to find the unsubtracted or $\overline{\text{MS}}$-subtracted results for $H^{d}_{\{\alpha,\beta,\gamma\}}$.

\section{The Mellin-Barnes Method \label{secMB}}

We give a brief overview of the basic Mellin-Barnes approach to Feynman integrals here. For a more comprehensive overview see \cite{Smirnov:2009up, Smirnov:2012gma, Friot:2011ic}. The Mellin transform is defined as follows:
\begin{align}
	[M(f)](s) = \int\limits_0^\infty f(t)t^{s-1}dt, \hspace{0.2in} s \in \mathcal{C}
\end{align}
Its inverse is given by:
\begin{align}
	[M^{-1} (g)](x) = \frac{1}{2\pi i} \int\limits_{c-i\infty}^{c+i\infty}  x^{-s}g(s)ds
\end{align}

The following formula derived from the inverse Mellin transform is used in high energy physics to write massive propagators as combinations of massless propagators:
\begin{align} \frac{1}{(m^2-k^2)^\lambda}=\frac{1}{2\pi i}\int\limits^{c+i \infty}_{c-i\infty} ds \frac{(m^2)^{-s}}{(-k^2)^{\lambda-s}}\frac{\Gamma(\lambda -s)\Gamma(s)}{ \Gamma (\lambda)} \label{MBformula}
\end{align}

The expression obtained after application of this formula and evaluation of the momentum integral is known as the Mellin-Barnes representation of a Feynman integral.

In some cases, it may be possible to simplify the Mellin-Barnes representation of an integral by the application of the following two Barnes lemmas \cite{Jantzen:2012cb}:
\begin{align}
& \frac{1}{2\pi i}\int_{-i\infty}^{i\infty} \Gamma(a+s)\Gamma(b+s)\Gamma(c-s)\Gamma(d-s)ds = \frac{\Gamma(a+c)\Gamma(a+d)\Gamma(b+c)\Gamma(b+d)}{\Gamma(a+b+c+d)} \\
\nonumber \\
& \qquad \text{and} \nonumber \\
\nonumber \\
&\frac{1}{2\pi i}\int_{-i\infty}^{i\infty} \frac{\Gamma(a+s)\Gamma(b+s)\Gamma(c+s)\Gamma(d-s)\Gamma(-s)}{\Gamma(e+s)}ds = \frac{\Gamma(a)\Gamma(b)\Gamma(c)\Gamma(d+a)\Gamma(d+b)\Gamma(d+c)}{\Gamma(e-a)\Gamma(e-b)\Gamma(e-c)}
\end{align}
where $ e \equiv a+b+c+d $

The evaluation of the Mellin-Barnes integrals may then be performed either numerically, or analytically by the addition of residues. In case of multiple Mellin-Barnes parameters, results from the theory of several complex variables may have to be used for analytic evaluation \cite{Friot:2011ic}.

\section{Derivative and Tensor Sunsets: $H'_{21}\{m_{\pi},m_K,m_K; m_{\pi}^2\}$ \label{secSpecialSunsets}}

In this section, we demonstrate how to handle both the tensor sunset integrals, as well as the derivatives of the sunsets, by reducing them to master integrals. In particular, we show how to evaluate the integral $H'_{21}\{m_{\pi},m_K,m_K;m_{\pi}^2\}$, by making extensive use of the package \texttt{Tarcer} \cite{Mertig:1998vk}. The computer implementation of what follows is given in the ancillary file \texttt{ReductionToMI.nb}.
The first step is to decompose $H_{21}\{m_{\pi},m_K,m_K; m_{\pi}^2\}$ into master integrals. From Eq.(\ref{H21}), we have:
\begin{align}
	H_{21} = \frac{\langle \langle (q.p)^2 \rangle \rangle d - \langle \langle q^2 \rangle \rangle s}{s^2(d-1)} 
\end{align}

Differentiating with respect to $s$ gives:
\begin{align}
	H'_{21} = \frac{ d \frac{\partial}{\partial s} \langle \langle (q.p)^2 \rangle \rangle - s \frac{\partial}{\partial s} \langle \langle q^2 \rangle \rangle + \langle \langle q^2 \rangle \rangle }{(d-1) s^2} - \frac{2 d \langle \langle (q.p)^2 \rangle \rangle}{(d-1) s^3} \label{H21prime}
\end{align}

The next step involves evaluating the scalar sunset integrals with $(q.p)^2$ and $q^2$ in the numerator. The following command allows us to express the first of these integrals in terms of the master integrals.\\

\texttt{TarcerRecurse[TFI[d, s, \{0, 0, 2, 0, 0\}, \{\{1, mpi\}, \{0, 0\},\{0, 0\},\{1, mk\},\{1, mk\}\}]]} \\

The output, $\langle \langle (q.p)^2 \rangle \rangle$, is a function of the dimensional parameter $d$, the external momentum $s$, the masses $m_{\pi}$ and $m_{K}$, the integrals $H^d_{\{1,1,1\}}\{m_{\pi},m_{K},m_{K};m_{\pi}^2\}$,  $H^d_{\{2,1,1\}}\{m_{\pi},m_{K},m_{K};m_{\pi}^2\}$,  $H^d_{\{1,1,2\}}\{m_{\pi},m_{K},m_{K};m_{\pi}^2\}$,  $A\{m_{\pi}\}$ and $A\{m_{K}\}$.

This expression is then differentiated with respect to $s$, the resulting expression, $\frac{\partial}{\partial s} \langle \langle (q.p)^2 \rangle \rangle$, also being a function of the same parameters and integrals as $\langle \langle (q.p)^2 \rangle \rangle$, but in addition also being a function of the differentiated master integrals $H'_{\{1,1,1\}}\{m_{\pi},m_{K},m_{K};m_{\pi}^2\}$,  $H'_{\{2,1,1\}}\{m_{\pi},m_{K},m_{K}; m_{\pi}^2\}$,  $H'_{\{1,1,2\}}\{m_{\pi},m_{K},m_{K}; m_{\pi}^2\}$.

Each of these differentiated master integrals can be expressed as a sunset integral in a higher ($d+2$) dimension by use of Eq.(\ref{DH}), and each of these higher dimensional sunsets can in turn be expressed in terms of the $d$ dimensional master integrals by further use of \texttt{Tarcer}. For example, the integral $H'_{\{2,1,1\}}\{m_{\pi},m_{K},m_{K};m_{\pi}^2\}\}$ is equal to $-2 H^{d+2}_{\{3,2,2\}}\{m_{\pi},m_{K},m_{K};m_{\pi}^2\}$. By use of the command:\\

\texttt{TarcerRecurse[TFI[d+2, s, \{\{3, mpi\}, \{0, 0\},\{0, 0\},\{2, mk\},\{2, mk\}\}]]}  \\

we get an expression for $H'_{2,1,1}\{m_{\pi},m_{K},m_{K};m_{\pi}^2\}\}$ in terms of $d$ dimensional master integrals. We repeat this process for each of the differentiated master integrals that appear, and substitute them (and $s=m_{\pi}^2$) into the expression for $\frac{\partial}{\partial s} \langle \langle (q.p)^2 \rangle \rangle$.

We can similarly obtain an expression for $\langle\langle q^2\rangle\rangle$ and $\frac{\partial}{\partial s}  \langle\langle q^2\rangle\rangle$, and substituting all these expressions into Eq.(\ref{H21prime}) with $s=m_{\pi}^2$ gives us our desired expression for $H'_{21}\{m_{\pi},m_K,m_K;m_{\pi}^2\}$.

The expressions we obtain for $H'_1$ and $H'_{21}$, given in the notebook \texttt{ReductionToMI.nb}, have been positively checked against expressions obtained from a direct differentiation of Eq.(2.13) and Eq.(2.14) of \cite{Kaiser:2007kf}, respectively.

\section{Single Mass Scale Sunset: $H^d_{\{1,1,1\}}\{ m_{\pi},m_{\pi},m_{\pi}; m_{\pi}^2 \}$ \label{sec1Mass}}

\subsection{Evaluation Using Mellin-Barnes}

All one mass scale sunset integrals can be reduced to a single master integral, namely $H^d_{\{1,1,1\}}\{ m,m,m; m^2 \}$ where $m$ is the mass in question. Below, we show how to evaluate the one mass scale sunset integral $H^{\chi}_{\{1,1,1\}}\{ m_{\pi},m_{\pi},m_{\pi};m_{\pi}^2 \}$, and therefore give a pedagogical demonstration of the use of the Mellin-Barnes approach to evaluating Feynman integrals. We also demonstrate the use of the public packages \cite{Gluza:2007rt} and \cite{Smirnov:2009up}. The accompanying Mathematica notebook \texttt{OneMassMB.nb} has a detailed computer implementation of what follows.

We begin by applying Eq.(\ref{MBformula}) to each of the propagators of the sunset integral Eq.(\ref{sunsetdef}) with $\alpha=\beta=\gamma=1$. We then combine a pair of (now massless) propagators by means of Feynman parameters, evaluate the integral over the loop momentum common to both propagators, and finally integrate over the Feynman parameter. This is then repeated with the result of the previous step and the remaining massless propagator to obtain the following Mellin-Barnes representation:
\begin{align}
	H^{\chi}_{\{1,1,1\}}&\{ m,m,m; m^2 \} = \nonumber \\
		& - \frac{(\mu_{\chi}^2)^{4-d}}{(4\pi)^d} \int \frac{\left(m^2\right)^{1-2 \epsilon} \Gamma(3-4 \epsilon-2 z) \Gamma(1-\epsilon-z)^2 \Gamma(-z)
\Gamma(\epsilon+z) \Gamma(-1+2 \epsilon+z)}{\Gamma(2-2 \epsilon-2 z) \Gamma(3-3 \epsilon-z)} dz \label{intMBdef}
\end{align}

To make contact with results in the literature, we extract a factor of $1/(4\pi)^d$. The above is also obtained automatically by use of the public code \cite{Gluza:2007rt}. The next step is to resolve (i.e separate) the singularities in $\epsilon$ and the finite part by shifting the contour across the points $z = 0$ and $z = 1-2\epsilon$. This can be done in an automatic manner by use of the package \cite{Smirnov:2009up}. The result is an expression consisting of two terms:
\begin{align}
	\frac{(4\pi)^d}{(\mu_{\chi}^2)^{4-d}} H^{\chi}_{\{1,1,1\}}&\{ m,m,m; m^2 \} = \nonumber \\
		& - \left(m^2\right)^{1-2 \epsilon} \Gamma(1-\epsilon) \Gamma(\epsilon)\Gamma(-1+2 \epsilon) \left(\frac{\Gamma(3-4 \epsilon) \Gamma(1-\epsilon)}{\Gamma(3-3 \epsilon)\Gamma(2-2 \epsilon)}+\frac{\Gamma(\epsilon)}{\Gamma(2-\epsilon)\Gamma(2 \epsilon)}\right)  \nonumber \\
		& -\int \frac{\left(m^2\right)^{1-2 \epsilon} \Gamma(3-4 \epsilon-2 z) \Gamma(1-\epsilon-z)^2
\Gamma(-z) \Gamma(\epsilon+z) \Gamma(-1+2 \epsilon+z)}{\Gamma(3-3 \epsilon-z) \Gamma(2-2\epsilon-2z)} dz
\end{align}

The first term contains the divergences, and the second piece is a finite one-fold contour integral which is to be evaluated by adding up residues. Since the singularities in $\epsilon$ have been extracted, we can set $\epsilon$ to 0 in the second term.

Expressing the divergent piece as a Laurent series around $\epsilon = 0$, we get:
\begin{align}
\frac{3 m^2}{2 \epsilon^2}&+\frac{m^2 \left(102-72 \gamma -72 \log\left(m^2\right)\right)}{24 \epsilon} \nonumber \\
& +\frac{m^2}{24}(201-204 \gamma + 72 \gamma^2 +14 \pi ^2-204 \log\left(m^2\right)+144 \gamma \log\left(m^2\right)+72
\log\left(m^2\right)^2) + \mathcal{O}(\epsilon)
\end{align}

The convergent piece is calculated by summing up the residues at the points $z = 0,1,2,3...$. The residues at non-zero integers $z=n+1$ for $n=0,1,2...$ are given by:
\begin{align}
2 m^2 \left(\frac{1}{n}+\frac{1}{1+n}+\frac{1}{2+n}\right)\frac{1}{n (1+n) (2+n)}
\end{align}
summing this up from $n = 1$ to $\infty$ gives:
\begin{align}
	\frac{3 m^2}{4}
\end{align}
The residue at $z_1$ = 0 is:
\begin{align}
	m^2 \left(-\frac{7}{4}-\frac{\pi ^2}{3}\right)
\end{align}

Combining the convergent and divergent pieces, we get the full result, expressed as a Laurent series in $\epsilon$:
\begin{align}
	\frac{(4\pi)^d}{(\mu_{\chi}^2)^{4-d}} H^{\chi}_{\{1,1,1\}}&\{ m,m,m; m^2 \} \nonumber \\
	& = \frac{3 m^2}{2 \epsilon^2}-\frac{m^2 \left(-17+12 \gamma+12 \log\left(m^2\right)\right)}{4 \epsilon}+\frac{1}{8}
m^2 (59+4 \gamma (-17+6 \gamma) +2 \pi ^2 \nonumber \\
	& + 4 \log\left(m^2\right) (-17+12 \gamma+6 \log\left(m^2\right)))
\end{align}

By pulling out a factor of $\Gamma(\epsilon)^2$ and setting $m$ to 1, this can be expressed more succinctly as:
\begin{align}
	H^{\chi}_{\{1,1,1\}} = \frac{(\mu_{\chi}^2)^{2\epsilon}}{(4 \pi)^{4-2\epsilon}} \left(\frac{3}{2}+\frac{17 \epsilon }{4}+\frac{59 \epsilon ^2}{8}\right) \Gamma(\epsilon)^2 \label{GS1mass}
\end{align}

This reproduces the result derived in Eq.(13) of \cite{Gasser:1998qt}. Expanding the above in powers of $\epsilon$, one gets the following result for the finite part of the $\overline{\text{MS}}_{\chi}$ subtracted single mass scale sunset integral:
\begin{align}
	H^{\chi}_{\{1,1,1\}}&\{ m,m,m;m^2 \} = \frac{m^2}{512 \pi^4} \bigg\{ 6 \log ^2\left(\frac{m^2}{\mu ^2}\right)-5 \log \left(\frac{m^2}{\mu ^2}\right)+\frac{\pi ^2}{2}+\frac{15}{4} \bigg\} \label{1massMB}
\end{align}

\subsection{Evaluation Using Tarcer}

The \texttt{Tarcer} package \cite{Mertig:1998vk} has the added functionality of performing a Laurent series expansion in the small parameter $\epsilon = (4-d)/2$ for the master integrals. The command for such an expansion is:

\begin{center}
	\texttt{TarcerExpand [Expression, $d \rightarrow 4- 2\epsilon$]}
\end{center}

\vspace*{3mm}

For one mass-scale sunsets, using this feature, \texttt{Tarcer} can be used directly to derive expressions for the integrals $H^d_{\{1,1,1\}}$, $H_{1}$, $H_{\mu\nu}$, $H'_{\{1,1,1\}}$, $H'_{1}$, $H'_{\mu\nu}$, i.e. for all the sunset results that appear in \cite{Gasser:1998qt}. This has been demonstrated in the notebook \texttt{OneMassTarcer.nb}, in which is derived a very comprehensive set of relations with detailed annotations, and completely verifies all the sunset relations in \cite{Gasser:1998qt}.

Note that the \texttt{TarcerExpand} command has been found to work for all the cases of interest, since this is a pure single mass scale example.  We find that for other more complicated mass configurations, including the case when we have a single mass scale with $s=0$, this command is unable to reproduce the Laurent expansion of the integral. However, that \texttt{Tarcer} can reproduce all the results for the sunsets in \cite{Gasser:1998qt} so efficiently indicates the power and utility of this package.

\section{Two Mass Scale Sunsets \label{sec2Mass}}

\subsection{Pseudothreshold Configurations: $H^{\chi}_{\{1,1,1\}} \{ m_{\pi}^2, m_K^2, m_K^2; m_{\pi}^2 \} $} \label{pt}

There are eight possible independent mass configurations of the sunset master integrals with two masses. Three of these fall into the pseudothreshold configurations, in which $s=(m_1+m_2-m_3)^2$. In the two-loop calculation of the pseudoscalar meson masses and decay constants, these are the only two-mass configurations that arise. Results for the pseudothresholds, calculated directly using an integral representation of the sunsets, are given in \cite{Berends:1997vk}. We rederived the three pseudothreshold results $H^d_{\{1,1,1\}}\{m,M,m;M^2\}$, $H^d_{\{2,1,1\}}\{m,M,M;m^2\}$ and $H^d_{\{1,2,1\}}\{m,M,M;m^2\}$ using Mellin-Barnes representations, and expressions for these are given below:

\begin{align}
H^{\chi}_{\{1,1,1\}}&\{m,m,M;M^2\} = \frac{M^2}{512 \pi ^4} \bigg[ -\log \left(\frac{m^2}{\mu ^2}\right)+2 \log ^2\left(\frac{M^2}{\mu ^2}\right)+2 \text{Li}_2\left(\frac{x}{x-1}\right) \nonumber \\
& -\log ^2\left(1-\frac{1}{x}\right)-\log ^2(x)-2 \log \left(\frac{x}{1-x}\right) \log \left(1-\frac{1}{x}\right)+\log (x)-\frac{1}{4}-\frac{\pi ^2}{6} \nonumber \\
& + 2 x \bigg( 2 \log ^2\left(\frac{m^2}{\mu ^2}\right)-2 \log \left(\frac{m^2}{\mu ^2}\right)-2 \text{Li}_2\left(\frac{x}{x-1}\right)+\log ^2(x-1)-\log ^2(x) \nonumber \\
& \qquad +2 \log \left(\frac{1}{1-x}\right) \log \left(\frac{x-1}{x}\right)-\log (x)+\frac{\pi ^2}{2}+2 \bigg) \nonumber \\
& + x^2 \bigg( 2 \text{Li}_2\left(\frac{x}{x-1}\right)-\log ^2(x-1)+\log ^2(x)-2 \log \left(\frac{1}{1-x}\right) \log \left(\frac{x-1}{x}\right)-\frac{\pi ^2}{3} \bigg) \bigg]
\end{align}

\begin{align}
H^{\chi}_{\{2,1,1\}}&\{m,M,M;m^2\} = \frac{1}{512 \pi ^4} \bigg[ 2 \log ^2\left(\frac{m^2}{\mu ^2}\right)+2 \log \left(\frac{m^2}{\mu ^2}\right)+\frac{\pi ^2}{3 x}-\log ^2(x)-\frac{\pi ^2}{6}-1 \nonumber \\
& + \left(1-\frac{1}{x}\right) \left(2 \text{Li}_2\left(\frac{1}{1-x}\right)+\log ^2(1-x)-2 i \pi  \log (1-x)\right) \bigg]
\end{align}

\begin{align}
H^{\chi}_{\{1,2,1\}}&\{m,M,M;m^2\} = \frac{1}{512 \pi ^4} \bigg[ 2 \log \left(\frac{m^2}{\mu ^2}\right)+2 \log ^2\left(\frac{M^2}{\mu ^2}\right)-\frac{\pi ^2}{3 x}+\frac{\pi ^2}{2}-1 \nonumber \\
& + \left(1-\frac{1}{x}\right) \left(-2 \text{Li}_2\left(\frac{1}{1-x}\right)-\log ^2(1-x)+2 i \pi  \log (1-x)\right) \bigg]
\end{align}
where $x=m^2/M^2$.

These results are valid for all real values of $x$. The other two mass pseudothreshold expressions may be obtained from the above by a simple re-ordering of the masses and indices. In the notebook \texttt{TwoMassPT.nb}, we demonstrate the above calculations by means of the example $H^{\chi}_{\{1,1,1\}} \{ m_{\pi}^2, m_K^2, m_K^2; m_{\pi}^2 \}$.

\subsection{Non-Pseudothreshold Configurations}

The evaluation of non-pseudothreshold two mass sunset configurations results in three complications that do not arise in the pseudothreshold case. Firstly, their Mellin-Barnes representation is a linear combination of complex-plane integrals of which at least one is two-fold, and which therefore requires a more sophisticated approach in its evaluation. These two-fold Mellin-Barnes integrals result in nested infinite sums, many of which cannot be expressed as common analytic functions. Therefore, completely analytic expressions for these integrals cannot be obtained easily, and we are forced instead to take as many terms of these sums as yields the degree of accuracy we desire. Secondly, the specific form of these infinite series depends on the numerical values of the two masses $m$ and $M$, or more specifically their ratio $m/M$. Thirdly, there exists a range of values of $m^2/M^2$ for which it is not possible to use the Mellin-Barnes method (given the current state of the art) to evaluate these integrals. For these values of $m^2/M^2$ one must make use of other techniques, such as expansion in the external momentum $s$.

The non-pseudothreshold mass configurations do not appear in the calculation of the pseudoscalar meson masses and decay constants to two-loops in chiral perturbation theory, but they may appear elsewhere. Thus for completeness we provide results for these as well in Appendix \ref{sec2MassNPT}. The notebook \texttt{TwoMassResults.nb} contains all the pseudothreshold and non-pseudothreshold two mass scale sunset integrals.

\section{Three Mass Scale Sunsets \label{sec3Mass}}

\subsection{Expansion in $s$: $H^{\chi}_{\{1,1,1\}} \left\{ m_{K}^2, m_{K}^2, m_{\eta}^2; m_{\pi}^2 \right\}$}

Three mass scale sunset integrals result in two-fold Mellin Barnes representations, which can be evaluated using 
the method of \cite{Friot:2011ic}. However, for purposes of evaluating the pion mass and decay constant, we take an expansion in the external momentum $s$:
\begin{align}
	H^{\chi}_{\{\alpha,\beta,\gamma\}} \{M,M,m;s\} &= H^{\chi}_{\{\alpha,\beta,\gamma\}}\{M,M,m;s=0\}  + s H'_{\{\alpha,\beta,\gamma\}}\{M,M,m;s=0\} \nonumber \\
	&  + \frac{s^2}{2!} H''_{\{\alpha,\beta,\gamma\}}\{M,M,m;s=0\} + \mathcal{O}(s^3)  
\end{align}

For the pion mass and decay constant the external momentum is always $s=m^{2}_{\pi}$, which is much smaller than the $m_K$ and $m_{\eta}$ that can appear in the propagators. Therefore, the above series converges fairly fast, and only a few of higher order terms are required. For integrals with $s=m_K^2$ or $s=m_{\eta}^2$, the Mellin-Barnes approach may be more suitable.

The derivatives of the integrals above can be evaluated using a combination of Eq.(\ref{DH}) and \texttt{Tarcer} \cite{Mertig:1998vk}. It turns out that derivatives to all orders of the sunset integral with $s=0$ can be expressed in terms of the single  master integral $H^{\chi}_{\{1,1,1\}}\{M,M,m;s=0\}$ given in Eq.(\ref{K111}).

\subsection{Two-Fold Mellin-Barnes Representations: $H^{\chi}_{\{1,1,1\}} \left\{ m_{K}^2, m_{K}^2, m_{\pi}^2; m_{\eta}^2 \right\} $}

For the three mass scale sunset integrals in which the external momentum is not the smallest parameter, such as those that appear in the kaon and eta masses and decay constants, the expansion in $s$ does not converge well. An expansion in one of the propagator masses must also be precluded as they lead to infrared divergences. The simplest method by which to obtain analytic expressions for these integrals to the order desired is by evaluating their two-fold Mellin-Barnes representation, a detailed explanation of which is given in \cite{Friot:2011ic}. In this section, we list the main intermediate results in the evaluation of $H^{\chi}_{\{1,1,1\}}\left\{ m_{K}^2, m_{K}^2, m_{\pi}^2; m_{\eta}^2 \right\}$ to exemplify the method in brief.

The first step is to find the Mellin-Barnes representation of the integral $H^{\chi}_{\{1,1,1\}}\left\{ m_{K}^2, m_{K}^2, m_{\pi}^2; m_{\eta}^2 \right\}$ and to resolve its singularity structure. This can be done semi-automatically by a combined use of the packages \texttt{AMBRE.m} and \texttt{MB.m}. The result is a linear combination of four parts. The first consists of the divergent parts and the finite part containing the $\mu$-scale dependent logarithms. The second and third parts are one-fold Mellin-Barnes integrals, the evaluation of which can be performed by simply adding up residues up to the desired order in powers of the mass ratio. The fourth part is proportional to the two-fold Mellin-Barnes representation:
\begin{align}
	 \int_{c-i \infty}^{c+i \infty} \int_{d-i \infty}^{d+i \infty} \frac{\Gamma^2 (1-z_1) \Gamma (2-z_1) \Gamma (-z_1) \Gamma (-z_2) \Gamma (z_1+z_2-1) \Gamma (z_1+z_2)}{\Gamma (2-2 z_1) \Gamma (z_2+2)} u_1^{z_1} (-u_2)^{z_2} \; dz_1 \; dz_2
\end{align}
where $u_1=m_K^2/m_{\pi}^2$, $u_2=m_{\eta}^2/m_{\pi}^2$, $c = 0.7$, $d=0.7$.

\begin{figure}[hbtp]
\centering
\includegraphics[scale=0.4]{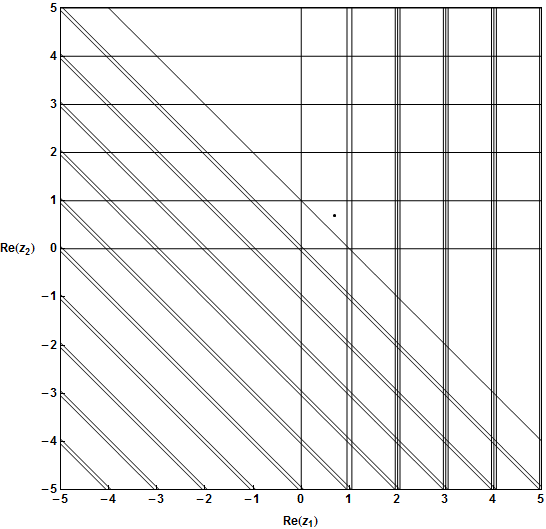}
\caption{Singularity Map of $H^{\chi}_{\{1,1,1\}}\left\{ m_{K}^2, m_{K}^2, m_{\pi}^2; m_{\eta}^2 \right\}$}
\label{SingMap}
\end{figure}

The singularity structure of this is given in Figure \ref{SingMap}. The poles whose residues are to be included in the summation are those at the intersection of the singularity lines.

\begin{figure}[hbtp]
\centering
\includegraphics[scale=0.5]{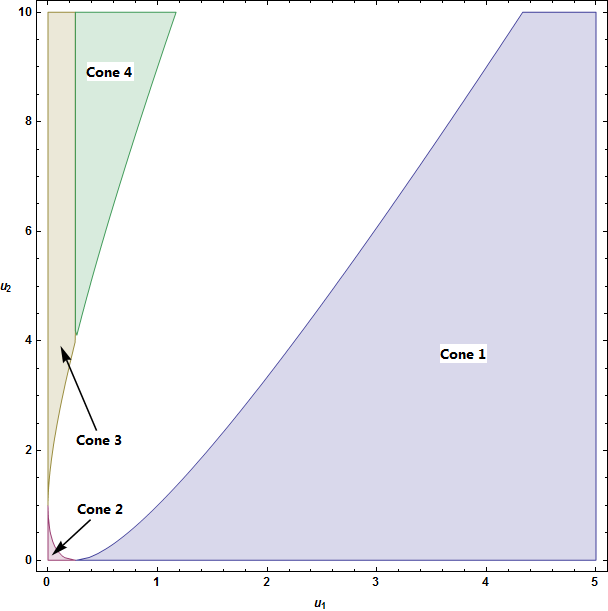}
\caption{Regions of convergence of $H^{\chi}_{\{1,1,1\}}\left\{ m_{K}^2, m_{K}^2, m_{\pi}^2; m_{\eta}^2 \right\}$}
\label{3MassConesFigure}
\end{figure}

The singularity structure above gives rise to four distinct cones, i.e. the above integral will converge to four distinct expressions depending on the particular value of the mass ratios $u_1$ and $u_2$. These regions are given in Table \ref{3MassConesTable} and plotted in Figure \ref{3MassConesFigure}.

\begin{table}
\centering
\begin{tabular}{|c|c|c|}
\hline 
\textbf{Cone} & \textbf{Region of Convergence} \\ 
\hline
Cone 1 & $1+\sqrt{|u_2|}<2\sqrt{|u_1|}$, $4|u_1|>|u_2|$, $4|u_1|>1$ \\ 
\hline 
Cone 2 & $|u_2|<1$, $4|u_1|<1$, $2\sqrt{|u_1|}+\sqrt{|u_2|}<1$ \\
\hline 
Cone 3 & $4|u_1|<|u_2|$, $|u_2|>1$, $4|u_1|<1$, $1+2\sqrt{|u_1|}<\sqrt{|u_2|}$ \\ 
\hline
Cone 4 & $4|u_1|<|u_2|$, $|u_2|>1$, $4|u_1|>1$, $1+2\sqrt{|u_1|}<\sqrt{|u_2|}$ \\ 
\hline
\end{tabular}
\caption{Regions of convergence of $H^{\chi}_{\{1,1,1\}}\left\{ m_{K}^2, m_{K}^2, m_{\pi}^2; m_{\eta}^2 \right\}$}
\label{3MassConesTable}
\end{table}

We see that there exists a large ``white space" which does not correspond to any of the four cones, i.e. it is not possible to directly use the Mellin-Barnes approach to derive an expression for the integral when the values of the mass-ratios $u_1$ and $u_2$ satisfy \{$1+\sqrt{|u_2|}>2\sqrt{|u_1|} \land 2\sqrt{|u_1|}+\sqrt{|u_2|}>1 \land 1+2\sqrt{|u_1|}>\sqrt{|u_2|}$\}.

To evaluate the two-fold integral above for cone 1 for example, we define the different singularity types that contribute to this cone by means of affine functions of $m$ and $n$:
\begin{align}
	& \text{Type 1} : \{z_1,z_2\} = \{0, 0\} \nonumber \\
	& \text{Type 2} : \{z_1,z_2\} = \{0, 1\} \nonumber \\
	& \text{Type 3} : \{z_1,z_2\} = \{0, -1\} \nonumber \\
	& \text{Type 4} : \{z_1,z_2\} = \{0, -2-m\} \nonumber \\
	& \text{Type 5} : \{z_1,z_2\} = \{-m-1, m+2\} \nonumber \\
	& \text{Type 6} : \{z_1,z_2\} = \{-m-n-1, m+1\} \qquad \text{where} \qquad m,n = 0,1,2,...
\end{align}

For each of these singularity types we shift the variables in the Mellin-Barnes representation by the affine functions to bring the poles to the origin. We then apply the reflection formula to all the gamma functions in the shifted representation that would be singular if evaluated with $z_1=0$ and $z_2=0$. This extracts the singularities to the denominator, from where they can be removed, and Cauchy's residue formula applied to the remaining integrand. (See \cite{Friot:2011ic} for more details.) This gives rise to a single residue, an infinite sum in $m$, or a double infinite series in $m$ and $n$, depending on the singularity type. For cone 1, we obtain (upto a factor of $m_{\pi}^2/256\pi^4)$:

\begin{align*}
\text{Type 1} = \frac{1}{2} \log ^2(-u_2)+\frac{\pi ^2}{6}+1
\end{align*}
\begin{align*}
\text{Type 2} = \frac{7}{4}u_2 + \frac{1}{2} u_2 \log \left( -\frac{u_1}{u_2} \right) 
\end{align*}
\begin{align*}
\text{Type 3} = \frac{1}{2 u_2}\log (-u_2) + \frac{5}{4u_2}
\end{align*}
\begin{align*}
\text{Type 4} &= - \frac{1}{u_2^2} \sum_{m=0}^{\infty}  \frac{\Gamma (m+1) \Gamma (m+2)}{\Gamma (m+3) \Gamma (m+4)} \left(\frac{1}{u_2}\right)^m \nonumber \\
 &=  \text{Li}_2\left(\frac{1}{u_2}\right)-\frac{1}{2} u_2 \log \left(1-\frac{1}{u_2}\right) +  \frac{1}{2 u_2} \log \left(1-\frac{1}{u_2}\right)-\frac{5}{4 u_2}  -\frac{1}{2}
\end{align*}
\begin{align*}
\text{Type 5} &= - \left( \frac{u_2^2}{u_1} \right) \sum_{m=0}^{\infty} \frac{\Gamma (m+1) \Gamma (m+2)^2}{\Gamma (m+4) \Gamma (2 m+4)} \left(\frac{u_2}{u_1}\right)^m = -\frac{u_2^2}{36 u_1}  \, _3F_2\left(1,1,2;\frac{5}{2},4;\frac{u_2}{4 u_1}\right)
\end{align*}
\begin{align}
\text{Type 6} = \left( \frac{u_2}{u_1} \right) & \sum_{m=0}^{\infty} \frac{\Gamma (m+n+1) \Gamma (m+n+2)^2 \Gamma (m+n+3)}{\Gamma (m+2) \Gamma (m+3) \Gamma (n+1) \Gamma (n+2) \Gamma (2 m+2 n+4)} \left(\frac{1}{u_1}\right)^n \left(\frac{u_2}{u_1}\right)^m  \nonumber \\
&  \times \bigg( \log (u_1) -\psi(m+n+1)-2 \psi(m+n+2)-\psi(m+n+3) \nonumber \\
& \quad +2 \psi(2 m+2 n+4) + \psi(n+1)+\psi(n+2) \bigg) \label{2foldresults}
\end{align}

Adding the results of the first three parts (those containing the $\mu$-dependent logarithms and those derived from the one-fold representations), as well as the contributions from Eq.(\ref{2foldresults}) up to the desired order gives us the analytic result for $H^{\chi}_{\{1,1,1\}} \left\{ m_{K}^2, m_{K}^2, m_{\pi}^2; m_{\eta}^2 \right\}$:

\begin{align}
H^{\chi}_{\{1,1,1\}}& \left\{ m_{K}^2, m_{K}^2, m_{\pi}^2; m_{\eta}^2 \right\} = \frac{m_{\pi}^2}{256\pi^4} \bigg[ \log ^2\left(\frac{m_{\pi}^2}{\mu ^2}\right)-\log \left(\frac{m_{\pi}^2}{\mu ^2}\right) -\frac{u_2^2}{36 u_1}  \, _3F_2\left(1,1,2;\frac{5}{2},4;\frac{u_2}{4 u_1}\right) \nonumber \\
& +\text{Li}_2\left(\frac{1}{u_2}\right)+\text{Li}_2(u_2) -\frac{1}{2} \log ^2(u_1)-2 \log (u_1)+\frac{1}{2} \log ^2(-u_2)+\frac{\pi ^2}{4}-\frac{5}{2} \nonumber \\
& +u_1 \left(2 \log ^2\left(\frac{m_{k}^2}{\mu ^2}\right)-2 \log \left(\frac{m_{k}^2}{\mu ^2}\right)+\frac{\pi ^2}{6}+3\right)+\frac{1}{8} u_2 \left(4 \log \left(\frac{m_{\pi}^2}{\mu ^2}\right)+4 \log (-u_1)+5\right) \nonumber \\
& + \left( \frac{1}{u_1} \right) \sum_{m=0}^{\infty} \frac{ \Gamma (m+1) \Gamma (m+2)}{\Gamma (2 m+4)} \left(\frac{1}{u_1}\right)^m  \bigg( \log (u_1)-\psi(m+1)-\psi(m+2)+2 \psi(2 m+4) \bigg) \nonumber \\
& + \left( \frac{u_2}{u_1} \right) \sum_{m,n=0}^{\infty} \frac{\Gamma (m+n+1) \Gamma (m+n+2)^2 \Gamma (m+n+3)}{\Gamma (m+2) \Gamma (m+3) \Gamma (n+1) \Gamma (n+2) \Gamma (2 m+2 n+4)} \left(\frac{u_2}{u_1}\right)^m \left(\frac{1}{u_1}\right)^n \nonumber \\
& \qquad \qquad \times \bigg( \log (u_1) -\psi(m+n+1)-2 \psi(m+n+2)-\psi(m+n+3) \nonumber \\
& \qquad \qquad \qquad +2 \psi(2 m+2 n+4) + \psi(n+1)+\psi(n+2) \bigg) \bigg] \label{3massMB}
\end{align}

The sums above can be evaluated to the desired order of the mass ratios. The order up to which the sums are required to be evaluated for a particular desired accuracy depend upon the numerical value of the mass-ratios. See Section \ref{secNumAnalysis} for a discussion of numerical issues.

\section{A One-Dimensional Representation for $H^d_{\{1,1,1\}}\{m,m,m;k m^2 \}$ \label{sec1DRep}}

For the sunset integral with the mass configuration $H^d_{\{1,1,1\}}\{m,m,m;k m^2\}$, which arises in SU(2) chiral perturbation theory, a Mellin-Barnes approach allows us an analytic expression that converges only for $k \geq 1$. Therefore, an alternative semi-analytic result is presented here for this mass configuration. The method used to derive the one-dimensional integral representation given in this section has been taken from the work of \cite{Berends:1997vk}.

By setting $m_1=m_2=m_3=m$ and applying the standard Feynman parametrization to Eq.(\ref{sunsetdef}), we get:
\begin{align}
H^d_{\{\alpha,\beta,\gamma\}}&\{m^2,m^2,m^2;s\} \nonumber \\
=& i^{-2d} (4\pi)^{-d} \frac{\Gamma(\alpha+\beta+\gamma-d)}{\Gamma(\alpha)\Gamma(\beta)\Gamma(\gamma)} \int^1_0 \int^1_0 \int^1_0 \frac{da_1 \; da_2 \; da_3}{(a_1a_2+a_1a_3+a_2a_3)^{\frac{3}{2}d-\alpha-\beta-\gamma}} \nonumber\\
& \qquad \qquad \frac{a_1^{\alpha-1}a_2^{\beta-1}a_3^{\gamma-1} \delta\left( \sum a_i-1 \right) }{\left( a_1a_2a_3s-(a_2a_3+a_1a_3+a_1a_2)(a_1+a_2+a_3)m^2 \right)^{\alpha+\beta+\gamma - d}}
\end{align}

By a series of algebraic manipulations we can rewrite the above integral as:
\begin{align}
H^d_{\{\alpha,\beta,\gamma\}}(km^2) & = i^{-2d} (4\pi)^{-d} \frac{\Gamma(\alpha+\beta+\gamma-d)}{\Gamma(\alpha)\Gamma(\beta)\Gamma(\gamma)} \int^1_0 \int^1_0 \int^1_0 da_1 \; da_2 \; da_3 \; \delta\left(\sum a_i - 1\right)  \nonumber \\ 
& \frac{a_1^{\beta+\gamma-\frac{d}{2}-1} a_2^{\alpha+\gamma-\frac{d}{2}-1} a_3^{\alpha+\beta-\frac{d}{2}-1}}{[(a_1 a_2 a_3 k - a_2 a_3 - a_1 a_3 - a_1 a_2)m^2]^{\alpha+\beta+\gamma - d}} \label{rep1}
\end{align}

Applying the Cheng-Wu theorem and rescaling the variables, we arrive at:
\begin{align}
H^d_{\{\alpha,\beta,\gamma\}}(km^2) &= i^{-2(\alpha+\beta+\gamma)} (4\pi)^{-d} (m^2)^{d-\alpha-\beta-\gamma}\frac{\Gamma(\alpha+\beta+\gamma-d)}{\Gamma(\alpha)\Gamma(\beta)\Gamma(\gamma)} \nonumber \\
& \times \int^\infty_0 \int^\infty_0 \frac{x^{\beta+\gamma-\frac{d}{2}-1}y^{\alpha+\gamma-\frac{d}{2}-1}dx dy}{(x+y+1)^{\frac{3d}{2}-\alpha-\beta-\gamma} \left[(x+y+1)(x+y+xy)-kxy\right]^{\alpha+\beta+\gamma-d}} \label{rep2}
\end{align}

Using Eq.(\ref{rep1}) and the relation:
\begin{align}
	\Gamma (-1+2\epsilon) = \frac {\Gamma (2\epsilon)}{-1+2\epsilon}
\end{align}
we can rewrite $H^{4-2\epsilon}_{\{1,1,1\}}$ as a linear combination of the integrals $ H^{6-2\epsilon}_{\{2,2,2\}} $, $H^{4-2\epsilon}_{\{2,1,1\}}$, $ H^{4-2\epsilon}_{\{1,2,1\}}$ and $H^{4-2\epsilon}_{\{1,1,2\}}$:
\begin{align}
H^{4-2\epsilon}_{\{1,1,1\}} & = \frac{m^2}{1-2\epsilon} \left( -k (4\pi)^{2} H^{6-2\epsilon}_{\{2,2,2\}} + H^{4-2\epsilon}_{\{2,1,1\}} + H^{4-2\epsilon}_{\{1,2,1\}}+H^{4-2\epsilon}_{1,1,2} \right) \nonumber \\
& = \frac{m^2}{1-2\epsilon} \left(-k(4\pi)^{2} H^{6-2\epsilon}_{\{2,2,2\}} + 3H^{4-2\epsilon}_{\{1,1,2\}} \right)
\end{align}

We can now compute the two integrals on the right hand side of the above relation using Eq.(\ref{rep2}). We begin our calculation with $H^{6-2\epsilon}_{\{2,2,2\}}$, first expanding the integrand around $\epsilon = 0$ up to $\mathcal{O}(\epsilon)$, and then integrating term by term to obtain the one-dimensional integral representation:
\begin{align}
	H^{6-2\epsilon}_{\{2,2,2\}}(km^2) &=m^{-4\epsilon}(4\pi)^{-6+2\epsilon}\nonumber \Gamma(2\epsilon) \\
	& \times \left(\frac{1}{2}+\frac{9\epsilon}{4}-2\epsilon \int_{0}^{\infty} \frac{s'}{(1+s')^3} \left[ -2+ \frac{2}{x'} \arctan(x') + \log(s') + \log(1+s') \right] ds' \right)
\end{align}
where 
\begin{align}
	x'(k,s') \equiv \sqrt{\frac{s'(s'+1-k)}{(k-5)s'-s'^2-4}} 
\end{align}

Note that $s'$ here is simply an integration variable, and is not related to the external momentum. To evaluate $H^{4-2\epsilon}_{\{1,1,2\}}$, we cannot directly expand the integrand in $\epsilon$ as it contains a divergent part. We first separate it into a divergent and a finite piece:
\begin{align}
	H^{4-2\epsilon}_{\{1,1,2\}} =  H_4^{div} + H_4^{fin}
\end{align}
and evaluate each piece separately. This gives:
\begin{align}
	H_4^{div}(km^2)=  (4\pi)^{-4+2\epsilon} \Gamma(2\epsilon)m^{-4\epsilon}\frac{\Gamma(2-4\epsilon)\Gamma^2(1-\epsilon)\Gamma(\epsilon)}{\Gamma(2-3\epsilon)\Gamma(2-2\epsilon)}
\end{align} and
\begin{align}
	H_4^{fin}(km^2) =  (4\pi)^{-4+2\epsilon}m^{-4\epsilon}\int_0^{\infty} \frac{s'}{(1+s')^2} \left[ -\frac{2}{x'} \arctan\left(x' \right) + \log\left( \frac{s'^2}{1+s} \right)  \right] ds'
\end{align}

Combining all the pieces produces the final one-dimensional integral representation up to $\mathcal{O}(\epsilon^2)$:
\begin{align*}
	H^{4-2\epsilon}_{\{1,1,1\}}&\{ m,m,m;km^2 \} \nonumber \\
	= & -\Gamma^2(\epsilon)(4\pi)^{-4+2\epsilon} m^2 \bigg\{ -\frac{3}{2} + \left[ -\frac{9}{2} + \frac{k}{4} + 3 \log(m^2) \right] \epsilon \nonumber \\
	& + \left[ -\frac{15}{2}+\frac{13 k}{8} - \pi^2 - \log(m^2) \left( \frac{k}{2} - 9 + 3 \log(m^2) \right)+\int_0^\infty f(k,s')ds' \right] \epsilon^2 \bigg\} 
\end{align*}
where
\begin{align}
	f(k,s') =&  -\frac{3s'}{(1+s')^2} \left[ -\frac{2}{x'} \arctan(x') + \log \left( \frac{s'^2}{1+s'} \right) \right] \nonumber \\
	& - \frac{ks'}{(1+s')^3} \left[ -2 + \frac{2}{x'} \arctan(x') + \log \left(s'\right) + \log \left(1+s' \right) \right]
\end{align}

We can rewrite this result in the following form to facilitate comparison with published results:
\begin{align}
	H^{4-2\epsilon}_{\{1,1,1\}}\{ m,m,m;km^2 \} &= \frac{(4\pi)^{-4+2\epsilon} m^{2-4\epsilon} \Gamma^2(1+\epsilon)}{(1-\epsilon)(1-2\epsilon)} \bigg\{ \frac{3}{2\epsilon^2}-\frac{k}{4\epsilon}-\frac{7k}{8}-3+\pi^2-\int_0^\infty f(k,s') \; ds' \bigg\}
\end{align}

Renormalizing the above using the $\overline{\text{MS}}_{\chi}$scheme, we obtain the result:
\begin{align}
	H^{\chi}_{\{1,1,1\}}&\{ m,m,m;km^2 \} \nonumber \\
		& =\frac{m^2}{512 \pi^4} \bigg\{ 3 + \frac{5 \pi^2}{2} - \frac{9 k}{4}  + 6 \log ^2\left(\frac{m^2}{\mu ^2}\right)+(k-6) \log \left(\frac{m^2}{\mu ^2}\right) - 2 \int_0^\infty f(k,s') \; ds' \bigg\} \label{OneDRep}
\end{align}

The only terms of $f(k,s')$ that are not analytically integrable are the ones containing the arctan factors. However, for the special values of $k = 1 \text{ and } 0$ an analytic integration of $f(k,s')$ is possible without any further substitutions. It may be possible to find a substitution for the case of $k=9$, for which the result is known exactly, but is beyond the scope of this discussion. When the integration is carried out, we produce the results given in Eq.(3.15) of \cite{Davydychev:1992mt} for $k=0$ and Eq.(27) of \cite{Berends:1997vk} for $k=1$. The case of $k=9$, which we have checked numerically, agrees with Eq.(28) of \cite{Berends:1997vk}.

For values of $x'$ with an imaginary part, the function $(2/x') \arctan(x')$ is pure real. The quadratic polynomial under the square root, $\sqrt{-4+(-5+k)s' - s'^2}$, determines whether $x'$ will be real or complex. For values of $k$ between 0 and 9, i.e. for values of the external momentum below the threshold, the quadratic polynomial has imaginary roots.

For values of $k > 9$, it is complex. An expression for the imaginary part for $k > 9$ is presented in Eq.(14) of \cite{Gasser:1998qt}, and in the the ancillary notebook \texttt{OneDRep.nb} we numerically demonstrate that the imaginary part generated by the integral representation Eq.(\ref{OneDRep}) agrees with this.

\section{Numerical Analysis \label{secNumerics}} 

In this section, we numerically compare the values obtained from the results given in Appendix \ref{sec2MassNPT} with those obtained by use of the program \texttt{Chiron} \cite{Bijnens:2014gsa} and \texttt{MB.m} \cite{Czakon:2005rk}.

\texttt{Chiron} is a code written in C++ for the express purpose of finding numerical values of the sunsets appearing in the meson masses and decay constants appearing in two loop SU(3) chiral perturbation theory. The \texttt{MBintegrate} function of \texttt{MB.m} \cite{Czakon:2005rk} is a more versatile tool that allows for the evaluation of non-sunset integrals as well from their Mellin-Barnes representations. However, while the scope of \texttt{Chiron} may be limited, within its range of applicability, a numerical comparison with previously published results shows \texttt{Chiron} to be highly accurate. Integrations performed using \texttt{MB.m} show variability in the accuracy of the results. A thorough study of the scope and limitations of \texttt{MB.m} remains to be done, but a first order examination shows that the accuracy of its results varies with the mass configuration and parameter values of the integral being evaluated. (See \cite{Gluza:2010rn}, however, for investigations into the efficiency of some aspects of these packages.)

The three mass scales that appear in chiral perturbation theory are the mass of the pion, kaon and eta, for which the latest values are given in \cite{Agashe:2014kda} as $m_{\pi} = m_{\pi^{\pm}} = 139.570$ MeV, $m_{K}=\sqrt{\tfrac{1}{2} (m_{K^+}^2 +m_{K^0}^2 - m_{\pi^+}^2 + m_{\pi^0}^2)} = 495.011$ MeV and $m_{\eta} = 547.862$ MeV. The following are the possible mass ratios with the above masses:

\begin{table}
\centering
\begin{tabular}{|c|c|}
\hline 
\bf{Mass ratio} $(x)$ & \bf{Numerical Value} \\ 
\hline 
$ m_{\pi}^2/m_{K}^2 $ & 0.07950 \\ 
\hline 
$ m_{\pi}^2/m_{\eta}^2 $ & 0.06490 \\ 
\hline 
$ m_{K}^2/m_{\pi}^2 $ & 12.57900 \\ 
\hline 
$ m_{K}^2/m_{\eta}^2 $ & 0.81637 \\ 
\hline 
$ m_{\eta}^2/m_{\pi}^2 $ & 15.40840 \\ 
\hline 
$ m_{\eta}^2/m_{K}^2 $ & 1.22493 \\ 
\hline
\end{tabular} \label{TableMassRatios}
\caption{List of all possible pseudoscalar meson mass ratios}
\end{table}

Using configuration 4, $H^{\chi}_{\{1,2,1\}}\{m,M,M;M^2\}$, as an example, we discuss issues concerning the speed of convergence and accuracy of the results given in Appendix \ref{sec2MassNPT} for the above given values of the mass ratios.

From Table \ref{TableMassRatios}, we see that the $\alpha$ series result given in Eq.(\ref{H121mMMalpha}) allows for the calculation of $H^{\chi}_{\{1,2,1\}}\{m,M,M;M^2\}$ for $x = m_{\pi}^2/m_{K}^2$, $m_{\pi}^2/m_{\eta}^2$ and $m_{K}^2/m_{\eta}^2$. The index $i$ in both the single and double sums of Eq.(\ref{H121mMMalpha}) controls the order of $x$, while the index $j$ in the double sum affects the accuracy of the result at a given order.

\begin{table}
\centering
\begin{tabular}{|c|c|c||c|c|c|} 
\hline 
\textbf{Mass Ratio} $(x)$ & \textbf{MB.m value} & \textbf{Asymptotic value} & $i$ & $j$ & \textbf{Min} $\mathcal{O}(x^n)$  \\
\hline 
$ m_{\pi}^2/m_{K}^2$ & $5.13510 \pm 0.00053 $ & $5.13509$ & $3$ & $5$ & $4$  \\ 
\hline 
$ m_{\pi}^2/m_{\eta}^2 $ & $4.79647 \pm 0.00053 $ & $4.79646$ & $3$ & $6$ & $4$ \\ 
\hline 
$ m_{K}^2/m_{\eta}^2 $ & $ 1.05418 \pm 0.00057 $ & $1.05418$ & $15$ & $5$ & $16$ \\ 
\hline
\end{tabular}
\label{TableH121alpha}
\caption{$H^{\chi}_{\{1,2,1\}}\{m,M,M;M^2\}$ calculated for three mass ratios that converge for the $\alpha$-series}
\end{table}

The second column of Table \ref{TableH121alpha} gives the value of the integral for the mass ratio given in the first column as computed using \texttt{MB.m}. The third column, labelled `Asymptotic value', gives the value of the integral as computed using Eq.(\ref{H121mMMalpha}) with the upper limit of the summation indices set to $i=j=500$. The next two columns give the lowest possible combination of values of the indices $i$ and $j$ which reproduce the asymptotic value. The order of $x$ this corresponds to is given in the last column, and is simply $n=i+1$. All numerical values in this table are given in units of $10^{-5}$.

The numbers in Table \ref{TableH121alpha} are indicative of general trends of all the $\alpha$-series results in Appendix \ref{sec2MassNPT}. As $ \lim_{x \to 1}$, the sums need to be taken a larger and larger order of $x$ to reach the asymptotic value. The minimum value of $j$ needed also generally increases with increasing $x$, but not necessarily. Furthermore, unless the summation is carried out to a sufficiently high order ($n$) of $x$, increasing solely the summation parameter $j$ tends the sum to a different limiting value from the actual value of the integral. Experimentation with the individual case at hand is necessary to determine the lowest values of $i$ and $j$ that yield the precision desired.

For the ratios $ m_{K}^2/m_{\pi}^2 $ and $ m_{\eta}^2/m_{\pi}^2 $, the $\beta$-series result Eq.(\ref{H121mMMbeta}) applies.

\begin{table}
\centering
\begin{tabular}{|c|c|c|c|c|c|} 
\hline 
\textbf{Mass Ratio} $(x)$ & \textbf{MB.m value} & \textbf{Asymptotic value} & \textbf{Min} $\mathcal{O}(x^n)$  \\
\hline 
$ m_{K}^2/m_{\pi}^2 $ & $ 1.81645 \pm 0.00013 $ & $1.81644$ & $ 16 $ \\ 
\hline 
$ m_{\eta}^2/m_{\pi}^2 $ & $ 1.60113 \pm 0.00014 $ & $ 1.60113 $ & $ 10 $ \\ 
\hline
\end{tabular}
\label{TableH121beta}
\caption{$H^{\chi}_{\{1,2,1\}}\{m,M,M;M^2\}$ calculated for two mass ratios that converge for the $\beta$-series}
\end{table}

In the case of the $\beta$-series results of Appendix \ref{sec2MassNPT}, both summation parameters $i$ and $j$ contribute to the order ($n$) of $x$, so a simpler correspondence between the value of $n$ and convergence can be made than in the case of the $\alpha$-series. Here too, the speed of convergence increases the further away from the lower possible bound of $x$ one is, i.e. convergence speeds up as $ \lim_{x \to \infty}$. The numbers in Table \ref{TableH121beta} are in units of $10^{-4}$.

An expression for $H^d_{\{1,2,1\}}\{m,M,M;M^2\}$ with mass ratio $ x = m_{\eta}^2/m_{K}^2 $ cannot be found using either of the two Mellin-Barnes derived series. An expansion in $s$ for this integral has been given as one possible means of dealing with this scenario. In Eq.(\ref{H121mMMsExp}) is given the expansion up to $\mathcal{O}(M^{10})$, but a numerical test up to $\mathcal{O}(M^{20})$ shows that the series tends to $1.72953 \times 10^{-6}$. The numerical result obtained for this integral from \texttt{MB.m} is $ (1.72961 \pm 0.006010) \times 10^{-6}$. Whether the series expansion converges accurately, or whether it converges to a value that is not the exact numerical value of the evaluated integral, cannot be determined at present due to the relatively large uncertainty accompanying the \texttt{MB.m} result.

\section{Conclusion and Discussion \label{secConcl}}

In this paper, we give a systematic account of how the different types and mass configurations of sunset diagrams appearing in SU(3) chiral perturbation theory may be analytically evaluated. In particular, we consider the reduction of vector and tensor sunsets to their scalar master integral constituents using integration by parts, and the evaluation of the sunset master integrals in which one, two and three different masses appear in the propagators or enter the loop as the external momentum squared. We use Mellin-Barnes representations in all these derivations, although other approaches (such as the differential equations method) have been successfully used previously to analytically evaluate some of the sunset configurations considered here. Our reason for preferring the Mellin-Barnes method was two-fold. Firstly, it expresses the results in an expansion of mass ratios, which is convenient for applications in an effective field theory such as chiral perturbation theory. Secondly, all the different mass configurations considered prove to be amenable to evaluation by use of a single method, i.e. the Mellin-Barnes representations, which therefore allows for a unified and consistent study of the subject.

In our evaluation of the sunsets, we make use of modern tools of the trade in the form of the publicly available packages \cite{Bijnens:2014gsa, Mertig:1998vk, Gluza:2007rt, Czakon:2005rk}. Indeed, one of the principal goals of this paper was to provide an analytical check on the results produced by these codes, and in particular \texttt{Chiron}, which as far as we are aware is the only package used for SU(3) chiral perturbation theory applications at two-loops. It must be pointed out that some of the codes listed above have capabilities far in excess of what was used in this paper, and future analytic work in this direction may require use of these capabilities. New versions of \texttt{Ambre.m} and \texttt{MBnumerics.m} \cite{Dubovyk:2016ocz}, for example, are capable of finding MB representations of non-planar diagrams, and evaluating them numerically to high precision.

We also provide as ancillary files to this work a set of Mathematica notebooks in which we demonstrate in greater detail the use of these packages in the evaluation of the sunsets. This allows the current paper to serve as a pedagogical introduction to the analytic evaluation of sunset integrals, as well as to the use of the available codes.

By way of original results, in Appendix \ref{sec2MassNPT} we present analytic expressions for all non-pseudothreshold two mass scale sunset integrals, which may be applicable in non-chiral perturbation theory contexts. These results are in the form of single and double infinite series, which converge for particular range of values of the mass ratio. The analytic continuation of these results to regions where the sums currently do not converge is currently under study. That Mellin-Barnes based calculations often lead to results that are not immediately convergent for input parameters over the whole complex plane is one of the major drawbacks of this approach. We also present an expansion in the external momentum for each of these integrals which allows one to obtain an analytic expression even for those values of the mass ratio for which the Mellin-Barnes derived results do not converge. The numerical analysis of Section \ref{secNumerics} shows that the Mellin-Barnes derived results converge fairly fast, and with excellent accuracy, for all values of the mass-ratio for which the result is valid. The speed of convergence and accuracy of the expansions in $s$, however, are dependent on the relative size of the two masses scale, and are generally not as reliable as the Mellin-Barnes derived results.

We also present an original one-dimensional integral representation of the sunset integral with one mass scale and arbitrary external momentum $H^d_{\{1,1,1\}}\{m,m,m;km^2\}$ that appears prominently in the context of SU(2) chiral perturbation theory. This representation can be evaluated fully analytically for $k=0 \text{ and } 1$, and can be evaluated semi-analytically for all other values of $k$. 

The novelty of the results presented in this paper lies in their analytic nature, which allows one to obtain numerical results of any desired degree of accuracy.

\section*{Acknowledgment}

It is a pleasure to thank Samuel Friot for explaining the nuances of the Mellin-Barnes method, Mikolaj Misiak for helpful comments on the manuscript, and Heinrich Leutwyler and Lorenzo Tancredi for helpful correspondence. JB is supported in part by the Swedish Research Council grants
contract numbers 621-2013-4287 and 2015-04089.

\appendix

\section{Non-Pseudothreshold Two Mass Scale Sunset Results \label{sec2MassNPT}}

\setcounter{equation}{0}
\renewcommand{\theequation}{A-\arabic{equation}}

The results for the pseudothreshold configurations are given in Section \ref{pt} of this paper. Here we list results for the other two-mass scale configurations. The range of values of $x = m^2/M^2 $ for which each of these expansions is valid is given in Table \ref{2massConv}. The expressions are generally not of a Horn's series type, which prevents one from computing the range of convergence using Horn's theorem. The entries of Table \ref{2massConv} have therefore been determined numerically.

\begin{table}
\centering
\begin{tabular}{|c|c|c|}
\hline 
\textbf{Integral} & \textbf{$\alpha$ series} & \textbf{$\beta$ series} \\ 
\hline 
$H_{\{1,1,1\}}\{m,M,M;M^2\}$ & $x < 1$ & $x \geq 8$ \\ 
\hline 
$H_{\{1,1,1\}}\{m,m,m;M^2\}$ & $x > 1$ & - \\ 
\hline 
$H_{\{2,1,1\}}\{m,M,M;M^2\}$ & $x < 1$ & $x \geq 9$ \\ 
\hline
$H_{\{2,1,1\}}\{M,m,M;M^2\}$ & $x < 1$ & $x \geq 9$ \\ 
\hline
$H_{\{2,1,1\}}\{m,m,m;M^2\}$ & $x > 1$ & - \\ 
\hline
\end{tabular}
\caption{Domain of convergence for the two mass scales sunset integral series of Appendix A}
\label{2massConv}
\end{table}

Also given for each mass configuration is the integral's expansion in $s$ up to a sufficient order in $s$, using which expressions may be derived for that range of $x$ not covered by either the $\alpha$ or $\beta$ series.

Both pseudothreshold and non-pseudothreshold results are also presented in the notebook \texttt{TwoMassScale.nb} for immediate computation. For notational convenience, we use the letter $K$ to refer to sunset diagrams with $s=0$ when writing out the integral as an expansion in the external momentum, i.e.
\begin{align*}
	K_{\{\alpha,\beta,\gamma\}} \{m_1,m_2,m_3 \} = H_{\{\alpha,\beta,\gamma\}} \{m_1,m_2,m_3; s = 0\}
\end{align*}

Also for notational convenience, we omit writing explicitly the mass configurations on the right hand side of the equations for the expansions in $s$, representing them using a bullet instead. For example,
\begin{align*}
H_{\{1,1,1\}} &\{m,M,M;M^2\} =  K_{\{1,1,1\}}\{\bullet\} + M^2 K'_{\{1,1,1\}}\{\bullet\} + \frac{M^4}{2!} K''_{\{1,1,1\}}\{\bullet\} + \mathcal{O}(M^6)
\end{align*}
is equivalent to
\begin{align*}
H_{\{1,1,1\}} \{m,M,M;M^2\} =&  K_{\{1,1,1\}}\{m,M,M \} + M^2 K'_{\{1,1,1\}}\{m,M,M \} + \frac{M^4}{2!} K''_{\{1,1,1\}}\{m,M,M \} \nonumber \\
& + \mathcal{O}(M^6)
\end{align*}

\vspace*{3mm}

\subsection*{Configuration 1: $H^{\chi}_{\{1,1,1\}}\{m,M,M;M^2\}$}

\subsubsection*{$\alpha$ series : $x < 1$}
\begin{align}
H_{\{1,1,1\}}^{\chi}&\{m,M,M;M^2\} = \frac{M^2}{512 \pi ^4} \bigg[ 4 \log ^2\left(\frac{M^2}{\mu ^2}\right)-3 \log \left(\frac{M^2}{\mu ^2}\right)+\frac{\pi ^2}{3}+\frac{7}{4}+\sqrt{3} \pi \nonumber \\
& + x \left(2 \log ^2\left(\frac{m^2}{\mu ^2}\right)-2 \log \left(\frac{m^2}{\mu ^2}\right)-\log ^2(x)+4 \log (x)+\frac{\pi ^2}{6}-5\right) \nonumber \\
& -2x^2 \sum_{i=0}^{\infty} \frac{\Gamma (i+1) \Gamma (i+2)}{\Gamma (2 i+4)} x^i \bigg( \psi(i+1)+\psi(i+2)-2 \psi(2 i+4)+\log (x) \bigg) \nonumber \\
& -2x \sum_{i,j=0}^{\infty} \frac{\Gamma (i+j+1) \Gamma (i+j+2)^2 \Gamma (i+j+3)}{\Gamma (i+2) \Gamma (i+3) \Gamma (j+1) \Gamma (j+2) \Gamma (2 i+2 j+4)} x^j  \nonumber \\
& \qquad  \times \bigg( \psi(i+j+1)+2 \psi(i+j+2)+\psi(i+j+3)-2 \psi(2 i+2 j+4) -\psi(j+1)-\psi(j+2)+\log (x) \bigg) \bigg]
\end{align}

\subsubsection*{$\beta$ series : $x \geq 8$}

\begin{align}
H_{\{1,1,1\}}^{\chi}&\{m,M,M;M^2\} = \frac{M^2}{512 \pi ^4} \bigg[ \log \left(\frac{m^2}{\mu ^2}\right)+4 \log ^2\left(\frac{M^2}{\mu ^2}\right)-4 \log \left(\frac{M^2}{\mu ^2}\right)-4 \text{Li}_2\left(\frac{x}{x-1}\right) \nonumber \\
& +4 \log \left(-\frac{1}{x^2}\right) \log \left(1-\frac{1}{x}\right)-6 \log ^2\left(1-\frac{1}{x}\right)-2 \log ^2(x)+\log \left(1-\frac{1}{x}\right)+\frac{\pi ^2}{3}-\frac{1}{4} \nonumber \\
& + x \bigg( 2 \log ^2\left(\frac{m^2}{\mu ^2}\right)-2 \log \left(\frac{m^2}{\mu ^2}\right)+4 \text{Li}_2\left(\frac{x}{x-1}\right)-2 \text{Li}_2\left(\frac{1}{x}\right)-4 \log \left(-\frac{1}{x^2}\right) \log \left(1-\frac{1}{x}\right) \nonumber \\
& \qquad +6 \log ^2\left(1-\frac{1}{x}\right)-\frac{\pi ^2}{6}+2 \bigg) -x^2 \log \left(1-\frac{1}{x}\right)  \nonumber \\
& - \frac{8}{\sqrt{\pi} x} \sum_{i=0}^{\infty} \frac{\Gamma \left(i+\frac{3}{2}\right)}{\Gamma (i+3)} \left(\frac{4}{x}\right)^{i} \bigg( \log^2\left(\frac{4}{x}\right) - \psi ^{(1)}(i+3)+\psi ^{(1)}\left(i+\frac{3}{2}\right)  \nonumber \\
& \qquad + \left(\psi\left(i+\frac{3}{2}\right)-\psi(i+3)\right)\bigg( -\psi(i+3)+\psi\left(i+\frac{3}{2}\right)+2 \log \left(\frac{4}{x}\right)\bigg) \bigg) \nonumber \\
& -\frac{8}{\sqrt{\pi } x^2} \sum_{i,j=0}^{\infty} \frac{\Gamma \left(i+\frac{3}{2}\right) \Gamma (i+j+2) \Gamma (i+j+3)}{\Gamma (i+1) \Gamma (i+2) \Gamma (i+3) \Gamma (j+2) \Gamma (j+3)} \left(\frac{4}{x}\right)^i \left(\frac{1}{x}\right)^j \nonumber \\
& \qquad \times \bigg( \bigg( -\psi(i+j+2)-\psi(i+j+3)+\psi(i+1)+2 \psi(i+2)+\psi(i+3)-2 \psi(2 i+3)+\log(x) \bigg)^2 \nonumber \\
& \qquad \quad + \psi ^{(1)}(i+j+2) +\psi ^{(1)}(i+j+3)-\psi ^{(1)}(i+1)-2 \psi ^{(1)}(i+2)-\psi ^{(1)}(i+3)+4 \psi ^{(1)}(2 i+3) \bigg) \bigg]
\end{align}

\subsubsection*{Expansion in $s$}
\begin{align}
H_{\{1,1,1\}}^{\chi} &\{m,M,M;M^2\} =  K^{\chi}_{\{1,1,1\}}\{\bullet\} + M^2 K'^{\chi}_{\{1,1,1\}}\{\bullet\} + \frac{M^4}{2!} K''^{\chi}_{\{1,1,1\}}\{\bullet\} + \mathcal{O}(M^6)
\end{align}
where
\begin{align}
K_{\{1,1,1\}}^{\chi} \{m,M,M\} = \frac{M^2}{512 \pi^4} &\bigg[ 4 \log ^2\left(\frac{M^2}{\mu ^2}\right)-4 \log \left(\frac{M^2}{\mu ^2}\right) + (x -4) F(x) +\frac{\pi ^2}{3}+6 \nonumber \\
& +  x  \left( 2 \log ^2\left(\frac{m^2}{\mu ^2}\right)-2 \log \left(\frac{m^2}{\mu ^2}\right)-\log ^2(x )+\frac{\pi ^2}{6}+3 \right) \bigg]
\end{align}
\begin{align}
K'^{\chi}_{\{1,1,1\}} \{m,M,M\} = \frac{1}{512 \pi^4 (x-4)^2} &\bigg[ x (x - 8) \log \left(\frac{m^2}{\mu^2}\right) + 16 \log \left(\frac{M^2}{\mu^2}\right) + \left(\frac{8}{x}-2\right) F(x) \nonumber \\
& +\left(\frac{x ^2}{4}-2 x +4\right) + 2 (x+4) \log (x) \bigg]
\end{align}
\begin{align}
K''^{\chi}_{\{1,1,1\}} \{m,M,M\} = \frac{1}{512 \pi^4 (x-4)^4 M^2} &\bigg[ \left(\frac{32}{x ^2}-4 x -\frac{40}{x }+24\right) F(x) + \left(\frac{4 x^2}{3}+\frac{4 x}{3}+\frac{32}{x}-\frac{104}{3}\right) \log (x) \nonumber \\
& +\left(-\frac{x ^3}{3}+\frac{10 x ^2}{3}-\frac{44 x }{3}-\frac{64}{x }+\frac{128}{3}\right) \bigg]
\end{align}

\vspace*{3mm}

\subsection*{Configuration 2: $H^{\chi}_{\{1,1,1\}}\{m,m,m;M^2\}$}

\subsubsection*{$\alpha$ series : $x>1$}
\begin{align}
H_{\{1,1,1\}}^{\chi}&\{m,m,m;M^2\} = \frac{M^2}{512 \pi ^4} \bigg[ \log \left(\frac{m^2}{\mu ^2}\right)+\frac{5}{4} -\frac{1}{18 x} \, _3F_2\left(1,1,2;\frac{5}{2},4;\frac{1}{4 x}\right) \nonumber \\
& + x \bigg( 6 \log ^2\left(\frac{m^2}{\mu ^2}\right)-6 \log \left(\frac{m^2}{\mu ^2}\right)+4 \sqrt{3} i \text{Li}_2\left( \frac{1}{4}+ \frac{\sqrt{3}}{4}i \right)-4 i \sqrt{3} \text{Li}_2\left( \frac{1}{4}- \frac{\sqrt{3}}{4}i \right) \nonumber \\
& \qquad +\frac{\pi ^2}{2}+9-\frac{4 \pi}{\sqrt{3}}\log (2) +\frac{2}{3}\psi ^{(1)}\left(\frac{1}{3}\right)-\frac{2}{3} \psi ^{(1)}\left(\frac{2}{3}\right) \bigg) \nonumber \\
& + 2 \sum_{i,j=0}^{\infty} \frac{\Gamma (i+j+1) \Gamma (i+j+2)^2 \Gamma (i+j+3)}{\Gamma (i+2) \Gamma (i+3) \Gamma (j+1) \Gamma (j+2) \Gamma (2 i+2 j+4)} \left( \frac{1}{x} \right)^i  \nonumber \\
& \qquad \times \bigg( -\psi (i+j+1) -2 \psi (i+j+2)-\psi (i+j+3)+2 \psi (2 i+2 j+4)+\psi (j+1)+\psi (j+2) \bigg) \bigg]
\end{align}

\subsubsection*{Expansion in $s$}
\begin{align}
H^{\chi}_{\{1,1,1\}}\{m,m,m;M^2\} &=  K^{\chi}_{\{1,1,1\}}\{\bullet\} + M^2 K'^{\chi}_{\{1,1,1\}}\{\bullet\} + \frac{M^4}{2!} K''^{\chi}_{\{1,1,1\}}\{\bullet\} + \frac{M^6}{3!} K'''^{\chi}_{\{1,1,1\}}\{\bullet\}  \nonumber \\ 
& + \frac{M^8}{4!} K''''_{\{1,1,1\}}\{\bullet\} + \mathcal{O}(M^{10})
\end{align}
where
\begin{align}
&  K^{\chi}_{\{1,1,1\}}\{m,m,m\} = \frac{m^2}{512 \pi^4} \bigg[ 6 \log ^2\left(\frac{m^2}{\mu ^2}\right)-6 \log \left(\frac{m^2}{\mu ^2}\right) -4\sqrt{3} \text{ Cl}_2 \left( \frac{\pi }{3} \right) + \frac{\pi ^2}{2}+9  \bigg]
\end{align}
\begin{align}
& K'^{\chi}_{\{1,1,1\}}\{m,m,m\} = \frac{1}{512 \pi^4} \bigg[ \log \left( \frac{m^2}{\mu ^2} \right) + \frac{8}{3\sqrt{3}} \text{ Cl}_2\left(\frac{\pi }{3}\right) + \frac{1}{4}  \bigg]
\end{align}
\begin{align}
& K''^{\chi}_{\{1,1,1\}}\{m,m,m\} = \frac{1}{512 \pi^4 m^2 } \bigg[ \frac{16}{27 \sqrt{3}} \text{ Cl}_2 \left( \frac{\pi }{3} \right) -\frac{11}{27} \bigg]
\end{align}
\begin{align}
& K'''^{\chi}_{\{1,1,1\}}\{m,m,m\} = \frac{1}{512 \pi^4 m^4} \left[ \frac{16 }{27 \sqrt{3}} \text{ Cl}_2 \left(\frac{\pi }{3}\right) -\frac{19}{54} \right]
\end{align}
\begin{align}
& K''''^{\chi}_{\{1,1,1\}}\{m,m,m\} = \frac{1}{512 \pi^4 m^6} \left[  \frac{256}{243 \sqrt{3}}\text{ Cl}_2 \left(\frac{\pi }{3}\right) - \frac{751}{1215} \right]
\end{align}

\vspace*{3mm}

\subsection*{Configuration 3: $H^{\chi}_{\{2,1,1\}}\{m,M,M;M^2\}$}

\subsubsection*{$\alpha$ series : $x < 1$}
\begin{align}
H^{\chi}_{\{2,1,1\}} & \{m,M,M;M^2\} = \frac{1}{512 \pi ^4} \bigg[ 2 \log ^2\left(\frac{m^2}{\mu ^2}\right)+2 \log \left(\frac{m^2}{\mu ^2}\right)-\log ^2\left(\frac{m^2}{M^2}\right)+2 \log \left(\frac{m^2}{M^2}\right) + \frac{\pi ^2}{6}-3 \nonumber \\
& - 2x \sum_{i=0}^{\infty} \frac{\Gamma (i+1) \Gamma (i+3)}{\Gamma (2 i+4)} x^i \bigg( \psi(i+1)+\psi(i+3)-2 \psi(2 i+4)+\log (x) \bigg) \nonumber \\
& - 2 \sum_{i,j=0}^{\infty} \frac{\Gamma (i+j+1) \Gamma (i+j+2)^2 \Gamma (i+j+3)}{\Gamma (i+1)^2 \Gamma (j+2) \Gamma (j+3) \Gamma (2 i+2 j+4)} x^i \nonumber \\
& \qquad \times \bigg( \psi (i+j+1)+2 \psi (i+j+2) +\psi (i+j+3)-2 \psi (2 i+2 j+4)-2 \psi (i+1)+\log (x) \bigg) \bigg]
\end{align}

\subsubsection*{$\beta$ series : $x \geq 9$}
\begin{align}
H^{\chi}_{\{2,1,1\}} & \{m,M,M;M^2\} = \frac{1}{512\pi^4} \bigg[ 2 \log ^2\left(\frac{m^2}{\mu ^2}\right)+2 \log \left(\frac{m^2}{\mu ^2}\right)+2 \text{Li}_2\left(\frac{1}{x}\right)+4 \log ^2\left(1-\frac{1}{x}\right) \nonumber \\
& -2 x \log \left(1-\frac{1}{x}\right)+4 \log (x) \log \left(1-\frac{1}{x}\right)+6 \log \left(1-\frac{1}{x}\right)+\frac{\pi ^2}{2}-1 \nonumber \\
& + \frac{2}{x^2} \sum_{i=0}^{\infty} \frac{\Gamma (2 i+3)}{\Gamma (i+1) \Gamma (i+3)} \left(\frac{1}{x}\right)^i \nonumber \\
& \qquad \times \left( (\psi(i+1)+\psi(i+3)-2 \psi(2 i+3)+\log (x))^2-\psi ^{(1)}(i+1)-\psi ^{(1)}(i+3)+4 \psi ^{(1)}(2 i+3) \right) \nonumber \\
& + \frac{2}{x^3} \sum_{i,j=0}^{\infty} \frac{\Gamma (2 i+3) \Gamma (i+j+3)^2}{\Gamma (i+1) \Gamma (i+2)^2 \Gamma (i+3) \Gamma (j+2) \Gamma (j+3)}\left(\frac{1}{x} \right)^{i+j} \nonumber \\
& \qquad \times \bigg( \bigg(\log (x)-2 \psi(i+j+3)+\psi(i+1)+2 \psi(i+2)+\psi(i+3)-2 \psi(2 i+3) \bigg)^2 \nonumber \\
& \qquad \qquad + 2 \psi ^{(1)}(i+j+3)-\psi ^{(1)}(i+1)-2 \psi ^{(1)}(i+2)-\psi ^{(1)}(i+3)+4 \psi ^{(1)}(2 i+3) \bigg) \bigg]
\end{align}

\subsubsection*{Expansion in $s$}
\begin{align}
H^{\chi}_{\{2,1,1\}}\{m,M,M;M^2\} &= K^{\chi}_{\{2,1,1\}}\{\bullet\} + M^2 K'^{\chi}_{\{2,1,1\}}\{\bullet\} + \frac{M^4}{2!} K''^{\chi}_{\{2,1,1\}}\{\bullet\} + \mathcal{O}(M^{6})
\end{align}
where
\begin{align}
K^{\chi}_{\{2,1,1\}} \{m,M,M\} = \frac{1}{512 \pi^4} \bigg[ 2 \log \left(\frac{m^2}{\mu^2}\right) + 2 \log^2 \left(\frac{m^2}{\mu^2}\right) + \left(1-\frac{2}{x}\right) F(x) -\log ^2(x )+\frac{\pi ^2}{6}+1 \bigg]
\end{align}
\begin{align}
K'^{\chi}_{\{2,1,1\}} \{m,M,M\} = \frac{1}{512 \pi^4 \left(x-4\right)^2 M^2} \bigg[ \frac{4}{x} \left(1-\frac{1}{x}\right) F(x)+\left(x +\frac{8}{x }-6\right)-\left(\frac{4}{x }+2\right) \log (x) \bigg]
\end{align}
\begin{align}
K''^{\chi}_{\{2,1,1\}} \{m,M,M\} = \frac{1}{512 \pi^4 \left(x-4\right)^4 M^4} & \bigg[ \left(-\frac{48}{x ^3}+\frac{56}{x ^2}-\frac{24}{x }+12\right) F(x) +\left(\frac{x ^2}{3} + \frac{96}{x ^2}-\frac{296}{3 x }+\frac{40}{3}\right) \nonumber \\
	& + \left(-\frac{48}{x ^2}-\frac{8 x }{3}+\frac{48}{x }-\frac{100}{3}\right) \log (x) \bigg]
\end{align}

\vspace*{3mm}

\subsection*{Configuration 4: $H^{\chi}_{\{1,2,1\}}\{m,M,M;M^2\}$}

\subsubsection*{$\alpha$ series : $x < 1$}
\begin{align}
	H^{\chi}_{\{1,2,1\}} & \{m,M,M;M^2\} = \frac{1}{512\pi^4} \bigg[ 2 \log ^2\left(\frac{M^2}{\mu ^2}\right)+2 \log \left(\frac{M^2}{\mu^2}\right) +\frac{\pi ^2}{6}+1 + \frac{1}{2} \, _3F_2\left(1,1,1;\frac{3}{2},3;\frac{1}{4}\right) \nonumber\\
	& + 2x \sum_{i=0}^{\infty} \frac{\Gamma (i+1) \Gamma (i+2)}{\Gamma (2 i+3)} x^i \bigg(\psi(i+1)+\psi(i+2)-2 \psi(2 i+3)+\log (x) \bigg) \nonumber \\
& + 2x \sum_{i,j=0}^{\infty} \frac{ \Gamma (i+j+2)^2 \Gamma (i+j+3)^2}{\Gamma (i+1) \Gamma (i+2) \Gamma (j+2) \Gamma (j+3) \Gamma (2 i+2 j+5)} x^i  \nonumber \\
& \qquad \times \bigg( \log (x) -\psi(i+1) -\psi(i+2)+ 2 \psi(i+j+2)+2 \psi(i+j+3)-2 \psi(2 i+2 j+5) \bigg) \bigg]
\label{H121mMMalpha}
\end{align}

\subsubsection*{$\beta$ series : $x \geq 9$}
\begin{align}
 H^{\chi}_{\{1,2,1\}} & \{m,M,M;M^2\} = \frac{1}{512\pi^4} \bigg[ 2 \log ^2\left(\frac{M^2}{\mu ^2}\right)+2 \log \left(\frac{M^2}{\mu ^2}\right)-\frac{\pi ^2}{6}+1-\log ^2(x) \nonumber \\
 & - \frac{2}{x} \sum_{i=0}^{\infty} \frac{\Gamma (i+1)}{\Gamma (i+3)} \left( \frac{1}{x} \right)^i \bigg( \psi(i+1)+\psi(i+2)-\log (x)+2 \gamma \bigg) \nonumber \\
 & - \frac{2}{x} \sum_{i=0}^{\infty} \frac{\Gamma (2 i+2)}{\Gamma (i+1) \Gamma (i+2)} \left( \frac{1}{x} \right)^i  \nonumber \\
 & \qquad \times \bigg( \bigg( \psi(i+1)+\psi(i+2)-2 \psi(2 i+2)+\log (x) \bigg)^2 - \psi ^{(1)}(i+1)-\psi ^{(1)}(i+2)+4 \psi ^{(1)}(2 i+2) \bigg) \nonumber \\
 & - \frac{2}{x^2} \sum_{i,j=0}^{\infty} \frac{\Gamma (2i+2) \Gamma (i+j+2) \Gamma (i+j+3)}{\Gamma (i+1)^2 \Gamma (i+2)^2 \Gamma (j+2) \Gamma (j+3)} \left(\frac{1}{x} \right)^{i+j} \nonumber \\
 & \qquad \times \bigg( \bigg( \psi(i+j+2)+\psi(i+j+3)-2 \psi(i+1)-2 \psi(i+2)+2 \psi(2 i+2)-\log (x) \bigg)^2 \nonumber \\
 & \qquad \quad + \psi ^{(1)}(i+j+2)+\psi ^{(1)}(i+j+3)-2 \psi ^{(1)}(i+1)-2 \psi ^{(1)}(i+2)+4 \psi ^{(1)}(2 i+2) \bigg) \bigg]
\label{H121mMMbeta}
\end{align}

\subsubsection*{Expansion in $s$}
\begin{align}
H^{\chi}_{\{1,2,1\}}\{m,M,M;M^2\} &=  K^{\chi}_{\{1,2,1\}}\{\bullet\} + M^2 K'^{\chi}_{\{1,2,1\}}\{\bullet\} + \frac{M^4}{2!} K''^{\chi}_{\{1,2,1\}}\{\bullet\} + \mathcal{O}(M^{6})
\label{H121mMMsExp}
\end{align}
where
\begin{align}
K^{\chi}_{\{1,2,1\}} \{m,M,M\} = \frac{1}{512 \pi^4 (x-4)} & \bigg[ -8 \log ^2\left(\frac{M^2}{\mu ^2}\right) -8 \log \left(\frac{M^2}{\mu ^2}\right) + (4-x) F(x) - \frac{2 \pi ^2}{3} - 4  \nonumber \\
& + 2 x \log ^2\left(\frac{M^2}{\mu ^2} \right)+ 2 x \log \left(\frac{M^2}{\mu^2}\right) + \left( \frac{\pi ^2 }{6}+ 1\right) x \bigg]
\end{align}
\begin{align}
K'^{\chi}_{\{1,2,1\}} \{m,M,M\} = \frac{1}{512 \pi^4 (x-4)^3 M^2} \bigg[ \left(-2 x -\frac{8}{x}+10\right) F(x) + \left(-x ^2+8 x -16\right) + ( x ^2-2 x -8 ) \log (x) \bigg]
\end{align}
\begin{align}
K''^{\chi}_{\{1,2,1\}}& \{m,M,M\} = \frac{1}{512 \pi^4 (x-4)^5 M^4 } \bigg[ \left(-4 x ^2-\frac{32}{x ^2}+16 x +\frac{40}{x }-8\right) F(x) \nonumber \\
& + \left( -\frac{5 x^3}{3}+\frac{22 x ^2}{3}+\frac{76 x }{3}+\frac{64}{x }-128 \right)  +\left(\frac{2 x ^3}{3}+\frac{40 x ^2}{3}-\frac{212 x }{3}-\frac{32}{x }+\frac{104}{3}\right) \log(x) \bigg]
\end{align}

\vspace*{3mm}

\subsection*{Configuration 5: $H^{\chi}_{\{2,1,1\}}\{m,m,m;M^2\}$}

\subsubsection*{$\alpha$ series : $x > 1$}
\begin{align}
	H^{\chi}_{\{2,1,1\}} & \{m,m,m;M^2\} = \frac{1}{512 \pi^4} \bigg[ 2 \log \left(\frac{m^2}{\mu ^2}\right) +2 \log ^2\left(\frac{m^2}{\mu ^2}\right)+ \frac{\pi ^2}{6}+1 + \frac{1}{2 x}\, _3F_2\left(1,1,1;\frac{3}{2},3;\frac{1}{4 x}\right) \nonumber \\
	& - 2 \sum_{i,j=0}^{\infty} \frac{\Gamma (i+j+1)^2 \Gamma (i+j+2)^2}{\Gamma (i+1) \Gamma (i+2) \Gamma (j+1) \Gamma (j+2) \Gamma (2 i+2 j+3)}
 \left( \frac{1}{x} \right)^i \nonumber \\
 & \qquad \times \bigg( \psi(j+1) +\psi(j+2) -2 \psi(i+j+1) -2 \psi(i+j+2)+2 \psi(2 i+2 j+3) \bigg) \bigg]
\end{align}

\subsubsection*{Expansion in $s$}
\begin{align}
H^{\chi}_{\{2,1,1\}}\{m,m,m;M^2\} &=  K^{\chi}_{\{2,1,1\}}\{\bullet\} + M^2 K'^{\chi}_{\{2,1,1\}}\{\bullet\} + \frac{M^4}{2!} K''^{\chi}_{\{2,1,1\}}\{\bullet\} + \frac{M^6}{3!} K'''^{\chi}_{\{2,1,1\}}\{\bullet\} \nonumber \\
& + \frac{M^8}{4!} K''''^{\chi}_{\{2,1,1\}}\{\bullet\} + \mathcal{O}(M^{10})
\end{align}
where
\begin{align}
K^{\chi}_{\{2,1,1\}}\{m,m,m\} = \frac{1}{512 \pi^4} \left[ \frac{1}{2} \left(2 \log \left(\frac{m^2}{\mu ^2}\right)+1\right)^2 -\frac{4}{\sqrt{3}} \text{Cl}_2\left(\frac{\pi }{3}\right) + \frac{1}{2} + \frac{\pi ^2}{6} \right]
\end{align}
\begin{align}
K'^{\chi}_{\{2,1,1\}}\{m,m,m\} = \frac{1}{512 \pi^4 m^2} \left( \frac{1}{3} \right)
\end{align}
\begin{align}
K''^{\chi}_{\{2,1,1\}}\{m,m,m\} = \frac{1}{512 \pi^4m^4 } \left[ \frac{11}{81}-\frac{16}{81 \sqrt{3}} \text{Cl}_2\left(\frac{\pi }{3}\right) \right]
\end{align}
\begin{align}
K'''^{\chi}_{\{2,1,1\}}\{m,m,m\} = \frac{1}{512 \pi^4 m^6 } \left[ \frac{19}{81}-\frac{32}{81 \sqrt{3}} \text{Cl}_2\left(\frac{\pi }{3}\right) \right]
\end{align}
\begin{align}
K''''^{\chi}_{\{2,1,1\}}\{m,m,m\} = \frac{1}{512 \pi^4 m^8 } \left[ \frac{751}{1215}-\frac{256}{243 \sqrt{3}} \text{Cl}_2\left(\frac{\pi }{3}\right) \right]
\end{align}

\vspace*{3mm}

\section{Public Codes Used in This Work \label{secPublicCodes}}

We present a brief description of each of the public packages referred to in this paper.\\

\texttt{Chiron} \cite{Bijnens:2014gsa} is a C++ program written to numerically evaluate the sunset diagrams that arise in the meson masses and decay constants at two-loop SU(3) chiral perturbation theory. It employs the notation of the mass and decay constant representations given in \cite{Amoros:1999dp}, and allows for a direct numerical evaluation of these quantities for variable mass input values. The results obtainable from \texttt{Chiron} are all in the $\overline{\text{MS}}_{\chi}$ scheme, and only the finite parts are presented. We use this code to check the results presented in this paper.\\

\texttt{Tarcer.m} \cite{Mertig:1998vk} is a Mathematica based package that automates the application of Tarasov's relations, i.e. it applies integration by parts to the input sunset diagram to express the integral as a linear combination of sunset master integrals and tadpole integrals. We make use of this package in the evaluation of the vector and tensor sunsets that appear in SU(3) chiral perturbation theory.\\

\texttt{AMBRE.m} \cite{Gluza:2007rt} is a Mathematica based package that takes as input any Feynman integral, and produces its Mellin-Barnes representation. It applies a loop-by-loop approach to the evaluation of the Mellin-Barnes representation, and thus produces representations that may not be the most efficient in terms of the number of Mellin-Barnes integrals. This can usually be reduced to its most efficient form, however, by application of the Barnes lemmas. This package was used extensively in this paper to obtain the Mellin-Barnes representation of the various sunset master integrals with differing mass configurations considered here. \\

\texttt{barnesroutines.m} \cite{Kosower} applies the first and second Barnes lemmas whenever possible to simplify a Mellin-Barnes representation.\\

\texttt{MBresolve.m} \cite{Smirnov:2009up} resolves the singularity structure of a given Mellin-Barnes integral using a different algorithm from the one used in MB.m. In our work, we primarily made use of this package for the resolution of the singularities, and used \texttt{MB.m} for the subsequent manipulations.\\

\texttt{MB.m} \cite{Czakon:2005rk} is another package written in Mathematica that takes as input a mellin-Barnes representation, and allows for various manipulations to be performed upon it. The functions of this program primarily used in this work were \texttt{MBexpand}, which allows for an expansion in $\epsilon$ to be taken, and \texttt{MBintegrate}, which numerically evaluates the Mellin-Barnes representation. We used this package extensively in the present work to expand in $\epsilon$ the singularity-resolved Mellin-Barnes representation from which the final analytic expressions were derived, as well as to numerically check our results.

\section{Notation and Dictionary \label{secNotation}}

\setcounter{equation}{0}
\renewcommand{\theequation}{C-\arabic{equation}}

In this section we provide a translation between the notation used in the calculation of \cite{ABG} and that used in the packages used for the calculation. \\

\texttt{AMBRE} defines its Feynman integrals as:
\begin{align}
	\int\int \frac{d^d k_1}{i \pi^{\frac{d}{2}}} \frac{d^d k_2}{i \pi^{\frac{d}{2}}} ...\frac{d^d k_n}{i \pi^{\frac{d}{2}}} \frac{X}{Y}
\end{align}

To account for the difference with Eq.(\ref{sunsetdef}), a factor of:
\begin{align}
	c1=\left( \frac{1}{4\pi} \right)^{4-2\epsilon}
\end{align}
needs to be multiplied to the sunset definitions in AMBRE. Another factor of:
\begin{align}
	c2=\left( \frac{\mu^2}{4\pi} \exp \left(\gamma_E - 1 \right) \right)^{2 \epsilon}
\end{align}
is also needed to introduce the $\overline{\text{MS}}_{\chi}$ subtraction to the sunset diagrams.

These may be introduced at the definition stage in \texttt{AMBRE.m} (i.e. by pre-multiplying the \texttt{Fullintegral} command). However, we found it more convenient to introduce these factors at the stage of expansion in $\epsilon$ after the residues had been resolved. Therefore, we introduced these factors when using the \texttt{MB.m} command \texttt{MBexpand}:

\begin{center}
	\texttt{MBexpand[mbrep, c1*c2, \{eps, 0, 0\}]}
\end{center}

The definition of the sunset diagram in the \texttt{Tarcer} package and Eq.(\ref{sunsetdef}) differs by an extra factor of $\left( \frac{1}{4\pi} \right)^{d}$ in the latter. Hence the need of the pre-factor in the following \texttt{Tarcer} definitions:
\begin{align}
 H(m_1,m_2,m_3,s)= -\frac{1}{(4\pi)^d} TFI[d, s, \{0,0,0,0,0\}, \{\{1,m_1\},\{0,1\},\{0,1\},\{1,m_3\},\{1,m_2\}\}] \nonumber \\
 p_\mu H^\mu(m_1,m_2,m_3,s) = -\frac{1}{(4\pi)^d} TFI[d, s, \{0,0,1,0,0\}, \{\{1,m_1\},\{0,1\},\{0,1\},\{1,m_3\},\{1,m_2\}\}] \nonumber \\
 p_\mu p_\nu H^{\mu\nu}(m_1,m_2,m_3,s) = -\frac{1}{(4\pi)^d} TFI[d, s, \{0,0,2,0,0\}, \{\{1,m_1\},\{0,1\},\{0,1\},\{1,m_3\},\{1,m_2\}\}] \nonumber \\
 g_{\mu\nu} H^{\mu\nu}(m_1,m_2,m_3,s) = -\frac{1}{(4\pi)^d} TFI[d, s, \{1,0,0,0,0\}, \{\{1,m_1\},\{0,1\},\{0,1\},\{1,m_3\},\{1,m_2\}\}]
\end{align}

Similarly, the integral:
\begin{align}
	A^{(d)}_{\{n,m\}} = \frac{1}{\pi^{d/2}}\int \frac{d^d k_1 } {[k_1^2-m^2]^{n}}
\end{align}

in \texttt{Tarcer} relates to the tadpole integral in Eq.(\ref{tadpoledef}) as:
\begin{align}
	A\{m\} = \frac{1}{(4 \pi)^{d/2}} \frac{1}{i} A^{(d)}_{\{1,m\}}  
\end{align}

The master integrals (with non-zero external momentum) in \texttt{Tarcer}:
\begin{align}
 J^{(d)}_{\{n_1,m_1\},\{n_2,m_2\},\{n_3,m_3\}} = \frac{1}{\pi^d}\int\int \frac{d^d k_1 d^d k_2 }{[k_1^2-m_1^2]^{n_1} [(k_1-k_2)^2-m_2^2]^{n_2}[k_2^2-m_3^2]^{n_3}  }
\end{align}

are related to $H^d_{\{1,1,1\}}$, $H^d_{\{2,1,1\}}$, etc. as:
\begin{align}
	H^d_{\{1,1,1\}}\{m_1,m_2,m_3;s\} = -\frac{1}{(4\pi)^d} J^{(d)}_{\{1,m_1\},\{1,m_2\},\{1,m_3\}} \nonumber \\
	H^d_{\{2,1,1\}}\{m_1,m_2,m_3;s\} = -\frac{1}{(4\pi)^d} J^{(d)}_{\{2,m_1\},\{1,m_2\},\{1,m_3\}} \nonumber \\
	H^d_{\{1,2,1\}}\{m_1,m_2,m_3;s\} = -\frac{1}{(4\pi)^d} J^{(d)}_{\{1,m_1\},\{1,m_2\},\{2,m_3\}} \nonumber \\
	H^d_{\{1,1,2\}}\{m_1,m_2,m_3;s\} = -\frac{1}{(4\pi)^d} J^{(d)}_{\{1,m_1\},\{2,m_2\},\{1,m_3\}} 
\end{align}

\vspace*{1cm}

\section{List of Ancillary Files} \label{secAncFiles}

We list the ancillary files provided with this work, and a brief description of their contents. \\

\begin{tabular}{|c|c|}
\hline 
\textbf{File }& \textbf{Description }\\ 
\hline
\texttt{ReductionToMI.nb} & Demonstrates how to use \texttt{Tarcer} to reduce all the varieties \\
& of sunset digrams to combinations of master integrals \\ 
\hline 
\texttt{OneMassMB.nb} & Demonstrates how to use a combination of AMBRE.m,\\
& MB.m, MRresolve.m and barnesroutines.m to evaluate sunsets \\ 
\hline
\texttt{OneMassTarcer.nb} & Demonstrates how to use \texttt{Tarcer} alone to derive all the one mass \\
& sunset diagrams required in ChPT calculations \\ 
\hline
\texttt{OneDRep.nb} & Presents a coded-in version of the one-dimensional representation\\& presented in Section 9, and checks its accuracy  \\ 
\hline
\texttt{TwoMassPT.nb} & Demonstrates the derivation of the integral $H^d_{\{1,1,1\}} \{ m_{\pi}^2, m_K^2, m_K^2; m_{\pi}^2 \} $ \\ 
\hline
\texttt{TwoMassResults.nb} & Contains results of all possible two mass scale configurations, \\ & both pseudothrehold and non-pseudothreshold \\
\hline 
\texttt{Miscellaneous.nb} & Contains expressions for several scalar, vector and tensor sunset \\
& integrals and their derivatives with $s=0$, as well as expressions\\& for the divergent part of the master integrals \\
\hline 
\end{tabular}\\

\end{document}